# Breaking symmetry to create a parallel-plate varactor dielectric with unparalleled microwave performance

Authors:
Florian Bergmann*[1,2], Matthew R. Barone*[3,4], Zishen Tian*[5,6], Aiden Ross*[7], Gerhard H. Olsen[8], Meagan C. Papac[1], Samuel Freed[1,9], Bryan T. Bosworth[1], Nicholas R. Jungwirth[1], Eric J. Marksz[1], Tomasz Karpisz[1,2], Noah Schnitzer[4,10], Lopa Bhatt[11], David A. Muller[10,11], Dylan Sotir[3,4], Akash Surampalli[5,6,12], Veronica Goian[13], Christelle Kadlec[13], Stanislav Kamba[13], Asher Hansen[14], Nathan Rongitsch[14], Dmitri A. Tenne[14], Ichiro Takeuchi[9], Long-Qing Chen[7], Lane W. Martin[5,12,15], Nathan D. Orloff[1], and Darrell G. Schlom[4,10,16]

* These authors contributed equally.

Affiliations:

[1] National Institute of Standards and Technology, Boulder, CO, USA

[2] Department of Physics, University of Colorado, Boulder, CO, USA

[3] Platform for the Accelerated Realization, Analysis, and Discovery of Interface Materials (PARADIM), Cornell University, Ithaca, NY, USA

[4] Department of Materials Science and Engineering, Cornell University, Ithaca, NY, USA

[5] Rice Advanced Materials Institute University, Rice University, Houston, TX, USA

[6] Department of Materials Science and Engineering, University of California, Berkeley, CA, USA

[7] Department of Materials Science and Engineering, Penn State University, State College, PA, USA

[8] Department of Materials Science and Engineering, NTNU – Norwegian University of Science and Technology, Trondheim, Norway

[9] Department of Material Science and Engineering, University of Maryland, College Park, MD, USA

[10] Kavli Institute at Cornell for Nanoscale Science, Cornell University, NY, USA

[11] School of Applied and Engineering Physics, Cornell University, NY, USA

[12] Department of Materials Science and NanoEngineering, Rice University, Houston, TX, USA

[13] Institute of Physics of the Czech Academy of Sciences, Prague, Czech Republic

[14] Department of Physics, Boise State University, Boise, ID, USA

[15] Departments of Chemistry and Physics and Astronomy, Rice University, Houston, TX, USA

[16] Leibniz-Institut für Kristallzüchtung, Berlin, Germany

## Abstract

Voltage-tunable capacitors (varactors) are key to microwave circuits. Tunable dielectric varactors outperform competing technologies in almost every relevant metric but usually suffer from high dielectric loss. In contrast, Ruddlesden-Popper (RPs) dielectric thin films have remarkably low microwave loss. Unfortunately, their crystallographic symmetry has until recently dictated an in-plane device structure, precluding the favorable out-of-plane parallel-plate varactor design for minimized size and maximized electric field in the tunable dielectric. Guided by theory, we report RPs akin to the widely studied tunable microwave dielectric $Ba_xSr_{1-x}TiO_3$. Assembling these same atoms into the first RP phase with broken out-of-plane symmetry, we achieve a low-loss, out-of-plane tunable dielectric thin film. The highest performing film, $(ATiO_3)_nAO$ film with $A$ = $Ba_{0.45}Sr_{0.55}$ and $n$ = 8, unlocks a tenfold improvement in the figure of merit for out-of-plane tunable dielectrics at 10 GHz, paving the way for a new generation of tunable monolithic microwave integrated circuits.

## Introduction

Varactors (tunable capacitors) are a fundamental design component for microwave electronics. They find use in tunable filters, impedance-matching networks, voltage-controlled oscillators, and phase shifters[1]. There are three major competing technologies for varactors: semiconductor diodes, micro-electromechanical systems, and tunable dielectrics. Semiconductor diode varactors dominate the low-frequency and -power market but suffer from limited power handling and high loss with increasing frequency[2,3]. Dielectric varactors outperform micro-electromechanical varactors in almost every metric – tuning range, operation voltage, switching speed, reliability, and price[4]. The only metric where micro-electromechanical varactors outperform typical dielectric varactors is their superior loss[3,4], motivating the search for low-loss tunable dielectrics.

Tunable dielectrics are (generally) ferroelectrics operated just above their ferroelectric – paraelectric phase transition temperature, but engineering highly tunable dielectrics with low loss is challenging. To date, some of the lowest-loss thin film tunable dielectrics are epitaxial Ruddlesden-Popper[5] (RP) titanates[6,7]: natural superlattices of perovskite, $A\mathrm{TiO_3}$, and rock salt, $A\mathrm{O}$, with formula $(ATi\mathrm{O_3})_nA\mathrm{O}$ where $n$ is the number of perovskite unit cells between rock-salt layers (**Fig. 1a-e**). These demonstrations utilized low loss RP phases, which are highly symmetric in bulk (I4/mmm), prohibiting ferroelectricity. Instead, their in-plane ferroelectricity relied on epitaxial strain breaking their in-plane symmetry, first yielding a high performance $(\mathrm{SrTiO_3})_6\mathrm{SrO}$ film[6], and subsequently, a $(\mathrm{SrTiO_3})_5(\mathrm{BaTiO_3})_1\mathrm{SrO}$ film with superior performance[7]. Until 2010, no inherently ferroelectric RP phases were known. In 2011, the discovery of hybrid improper ferroelectricity (HIF)[8] transformed the landscape of ferroelectrics. With it came a wave of HIF RPs that were thought to be[9] and subsequently shown to be inherently ferroelectric[10–12]. Until recently, all known (predicted or experimentally established) ferroelectric RPs had spontaneous polarizations parallel to the rock-salt layers[13] and in the plane of epitaxial thin films. While never demonstrated epitaxially, a layered crystal akin to RPs — $Li_2(Sr_{1-x}Ca_x)Nb_2O_7$ — recently displayed spontaneous polarization (space group $Pna2_1$, where the polar $c$ axis is perpendicular to the rock-salt layers)[14] and a Curie temperature that could be tuned from 217 K to above ambient temperature by increasing its calcium content[15]. Later, theorists predicted several RPs with out-of-plane spontaneous polarization based sulfides and halides (*e.g.*, $SrTb_2Fe_2S_7$ and $KAg_2Mn_2Cl_7$,[16]), but none of these have been demonstrated nor implemented in the commercially relevant varactor geometries based on metal-insulator-metal (MIM) capacitors (*i.e.*, parallel-plate capacitors with an out-of-plane electric field). Compared to in-plane varactors, which initially revealed low loss in RP thin films[6,7], MIM varactors require a lower tuning voltage, have a smaller spatial footprint, and maximize tunability by confining the electric field within the tunable dielectric.

Equivalently to previous work breaking in-plane symmetry to achieve an in-plane ferroelectric RP, we now aim to break the out-of-plane symmetry to achieve an out-of-plane ferroelectric RP phase for MIM varactors. Our own first-principles calculations indicate that it might be possible to create an oxide RP phase akin to one of the most widely studied tunable microwave dielectrics, $Ba_xSr_{1-x}TiO_3$, with monoclinic symmetry (space group $Cm$)[17], which hosts out-of-plane ferroelectricity. Inspired by these calculations, we construct and characterize parallel-plate varactors containing $(ATiO_3)_nAO$ with $A$ = $Ba_{0.45}Sr_{0.55}$ and different rock-salt-layer periodicities $n$. These RP phases all display out-of-plane spontaneous polarization at low temperatures and high tunability at ambient temperature. The top performing dielectric, $(ATiO_3)_8AO$, surpasses the 10 GHz dielectric tuning figure of merit of the best $Ba_xSr_{1-x}TiO_3$ out-of-plane tunable dielectric ever reported by an order of magnitude.

## Finding a ferroelectric RP phase

We investigate the possibility of a paraelectric-to-ferroelectric phase transition in $(A\mathrm{TiO_3})_nA\mathrm{O}$, $A$ = $\mathrm{Ba}_x\mathrm{Sr}_{1-x}$, with zero-temperature density functional theory (**Methods**). We simulate the structure with different periodicity of rock-salt layers ($n$ = 8, 12, 16, 20, and ∞) and barium content ($x$ = 0.3, 0.5, and 0.7). The periodicity $n = \infty$ corresponds to the pure $\mathrm{Ba}_x\mathrm{Sr}_{1-x}\mathrm{TiO_3}$ perovskite phase with no rock-salt layers. Starting from the non-polar structure with space group symmetry *I*4/*mmm*, we extract the squared frequency of the softest polar phonon (**Figs. 1f and S1**). A positive squared frequency indicates that the non-polar structure is stable, while a negative squared frequency indicates an instability that distorts the crystal lattice towards a polar space group. The phonon calculations show in-plane polarization for essentially all periodicities and barium contents with out-of-plane polarization emerging only for sufficiently high values of $n$ and $x$ (*i.e.*, when the composition becomes increasingly akin to that of the perovskite $\mathrm{BaTiO_3}$). This said, the calculations predict that $(A\mathrm{TiO_3})_nA\mathrm{O}$ films with $n \geq 16$ have both an in-plane and an out-of-plane polarization component (space group *Cm*).

To investigate the impact of the rock-salt-layers on the dielectric tunability at ambient temperature, we perform dynamical phase-field simulations (**Methods**). The simulations indicate that $(A\mathrm{TiO_3})_nA\mathrm{O}$, $A$ = $\mathrm{Ba_{0.45}Sr_{0.55}}$ retains out-of-plane dielectric tunability despite the presence of the rock-salt layers, although the dielectric tunability decreases as the periodicity $n$ decreases (**Fig. 1g**). At the same time, the tuning peak width $E_0$ monotonically increases for decreasing periodicity $n$ (**Fig. 1g, right inset**). The existence of an out-of-plane polarization in the first-principles calculations and the persistent out-of-plane dielectric tunability in the phase-field simulations suggests that we can transfer loss-mitigation strategies previously utilized for low-loss, in-plane tunable dielectrics to mitigate the dielectric loss in out-of-plane tunable dielectrics.

## Growing RPs with high barium content and dilute rock-salt layers

We use molecular-beam epitaxy (MBE, **Methods**) to synthesize four different variants of MIM heterostructures containing 200 nm $(A\text{TiO}_3)_nA\text{O}$ dielectrics with $A$ = $Ba_{0.45}Sr_{0.55}$ and $n$ = 8, 16, 24, and ∞. Such RP titanates are metastable; the stability limit in bulk is limited to barium contents $x$ < 0.03 at 1400 °C[18]. Nonetheless, we synthesize these RPs with $x$ = 0.45 by MBE. In bulk, the (stable) perovskite composition $Ba_{0.45}Sr_{0.55}TiO_3$ transitions to the ferroelectric phase near 200 K[19], similar to the highest performing tunable dielectrics observed to date[6,7]. Scanning transmission electron microscopy (STEM), electron energy loss spectroscopy (EELs) (**Figs. 1a-c**), and X-ray diffraction (XRD, **Figs. 2a**) reveal clear ordering of rock-salt layers with no evidence of secondary phases. Reciprocal space maps (RSMs) confirm that all layers of the heterostructures are commensurately strained to the (110)-oriented $DyScO_3$ substrate (**Fig. 2b**). The dielectric is almost unstrained (-0.1 % biaxial compression). The STEM image (**Fig. 1a-c**) of the heterostructure containing $(A\text{TiO}_3)_{16}A\text{O}$ depicts the intended horizontal $(A\text{O})_2$ rock-salt layers separated by 16 perovskite unit cells consistent with the XRD data (**Fig. 2a**). In addition, we observe regions where the intended horizontal rock-salt layers fail to crystallize, which causes local formation of higher-order phases (*e.g.*, $n$ = 32). Wherever horizontal rock-salt layers are absent, vertical $(A\text{O})_2$ planes naturally crystallize to avoid forming high-energy dislocations[20]. These junctions also cause stresses that likely influence the local dielectric response[21]. All low-magnification STEM images of RP films with $n$ > 3 reveal similar defects[6,7,22–28], and previous works have argued that these *uncharged* defects are the key to low dielectric loss in RP phases[6,7]. The prominence of XRD peaks in the RP films scales with the successful crystallization of horizontal rock-salt layers. We suspect that the higher intensity of XRD peaks in the lower $n$ films indicates more complete crystallization of horizontal $(A\text{O})_2$, but it is not clear to what extent the different crystal structure influences relative intensities. While XRD data cannot directly speak to the magnitude of the point-defect concentration, the clarity of the Laue oscillations correlates with the long-range order in the crystal. Clear Laue oscillations from both the 200-nm-thick $Ba_xSr_{1-x}TiO_3$ and the $SrRuO_3$ electrodes support its use as a high-quality control sample (**Fig. 2a, inset**). Having demonstrated synthesis of these metastable RPs, it remains to validate our prediction that these phases will display out-of-plane ferroelectricity, and if so, assess their microwave performance against our $Ba_xSr_{1-x}TiO_3$ control sample.

## Confirming out-of-plane ferroelectricity at low temperatures

We characterize the temperature-dependent permittivity (**Fig. 3a, top**) at 1 kHz, 10 kHz, and 100 kHz with MIM capacitors that we grounded via an exposed pad of the $SrRuO_3$ bottom electrode (**Methods**, **Fig. S11**). We observe a maximum in the temperature-dependent permittivity, indicating a paraelectric-to-ferroelectric phase transition in all samples. All dielectric-maximum temperatures $T_{\mathrm{m}}$ fall in a narrow range (183 K < $T_{\mathrm{m}}$ < 253 K). On the other hand, the extracted Curie-Weiss temperature $T_0$ monotonically decreases as the periodicity $n$ decreases to the point that it ceases to exist for the $n$ = 8 film (**Fig. 3a, bottom**), exactly as predicted by the first-principles calculations (**Fig. 1f**). Only for the $n$ = ∞ perovskite do we extract $T_0$ ≈ $T_{\mathrm{m}}$; the RP films exhibit $T_{\mathrm{m}} > T_0$. Furthermore, the $n$ = ∞ perovskite films have the sharpest phase transition, while a decreasing periodicity $n$ broadens the transition (**Fig. 3b**). Interestingly, the $n$ = 24 films have the highest $T_{\mathrm{m}}$ (**Fig. 3b, inset**).

We look to polarization-electric field (*P*-*E*) loops to interpret these trends. At 30 K, all films exhibit out-of-plane ferroelectric(-like) hysteresis (**Fig. 3c**) and this hysteresis persists up to temperatures significantly above $T_0$ and closer to $T_{\mathrm{m}}$ (135 K for $n$ = 8 film, and 285 K for $n$ = 24 film, **Fig. S15**). At 300 K, all films exhibit paraelectric behavior (**Fig. 3c, inset**). These observations suggest that $T_{\mathrm{m}}$ represents the ferroelectric-to-paraelectric transition temperature rather than $T_0$. Additional characterization of the transition temperature with UV Raman spectroscopy (**Fig. S9**) and THz and IR spectroscopy (**Fig. S7**) support this interpretation. We resolve the apparent contradiction of different $T_{\mathrm{m}}$ and $T_0$ by recalling that we can understand $T_0$ as a projected temperature for soft-phonon-mode condensation[29]. As such, $T_0$ represents the "intrinsic" ferroelectric instability that the first-principles calculations predict. In contrast, "extrinsic" factors such as defects or interfaces affect $T_m$. This view suggests that such extrinsic factors define the phase transition in the RP films. The broadening of the transition with decreasing periodicity $n$ underpins this interpretation.

In total, the cryogenic characterization illustrates that all RP films have an out-of-plane ferroelectric-to-paraelectric phase transition, confirming the theoretical predictions that $(A\mathrm{TiO_3})_n A\mathrm{O}$ with $A$ = $Ba_{0.45}Sr_{0.55}$ breaks out-of-plane symmetry. The characterization also reveals that all films exhibit out-of-plane tunability at ambient temperature (**Fig. 3d**). The unbiased permittivity confirms the qualitative prediction from the phase-field simulations ($\varepsilon_{\mathrm{measured}} \approx 0.7\ \varepsilon_{\mathrm{simulated}}$, **Fig. 3d inset**). Interestingly, the dielectric loss is highest for the $n$ = 8 film and lowest for the $n$ = ∞ film (**Fig. 3d**). This trend slightly dims our hope that we could use RPs to engineer an out-of-plane tunable low-loss microwave dielectric but as we see next, this turns out not to be the case.

## Observing low microwave loss and high tunability

To measure the frequency- and voltage-dependent complex out-of-plane permittivity at microwave frequencies, we rely on the recently updated metrology for out-of-plane permittivity extraction (**Methods** and Ref. [30]). It uses structures that are compatible with ground-signal-ground probes, incorporates an improved calibration technique and accounts for inhomogeneous field patterns. While the real permittivity between 10 MHz to 300 MHz agrees with the values extracted at 10 kHz, above 300 MHz, the films show distinct frequency dispersions that follow the Havriliak-Negami model[31] (**Fig. 4a,b**). The ability to describe the complex permittivity dispersion with the Havriliak-Negami model suggest that a single polarization mechanism dominates the frequency dispersion in each film. The rock-salt layers clearly influence this polarization mechanism and lead to reduced dispersion and microwave loss. The critical periodicity to achieve loss suppression is between 8 < $n$ < 16, with the most significant increase in dispersion occurring at 16 < $n$ < 24. This critical periodicity also ensures a substantial increase of the spatial homogeneity in the film permittivity above 300 MHz across devices at different locations on a 10 mm × 10 mm chip (**Fig. S20**), showing that the RP phase also results in superior spatial homogeneity. This superior homogeneity is a crucial requirement for reproducible material and device properties and a critical challenge in ferroelectrics to date[32]. While the tuning curves in the range from 10 MHz to 300 MHz (**Fig. 4c, left inset**) are comparable to the ones at 10 kHz for all films (**Fig. 3d**), the tunings of the $n$ = ∞ and $n$ = 24 films at 10 GHz are drastically reduced due to their strong dispersion but the $n$ = 16 and $n$ = 8 maintain their tunability (**Fig. 4c**).

While the measured tuning peak width of the $n$ = ∞ perovskite at 0.1 GHz ($E_0$ = 74 kV/cm) is comparable to the value predicted by phase-field simulations ($E_0$ = 92 kV/cm), the tuning peak (**Figs. 4c, right inset**) does not broaden with decreasing periodicity $n$ as predicted (**Fig. 1d, inset**). Instead, the $n$ = 24 film has the sharpest tuning curve, and the peak width as a function of the periodicity $n$ rather resembles the trend of $T_{\mathrm{m}}$ (**Fig. 3c, inset**). This correlation suggests that it is the vicinity to $T_{\mathrm{m}}$ that is relevant for the tunability at ambient temperature. As discussed in the previous section, extrinsic effects dictate the transition temperature $T_{\mathrm{m}}$, and extrinsic effects like local stress fields near the vertical and horizontal rock-salt layers could cause stress-induced polar nanoregions[21] that are unaccounted for in the phase-field models.

The mechanism of loss suppression in RP titanates is still under debate. Earlier microwave experiments attributed the dispersion in $Ba_xSr_{1-x}TiO_3$ to charged point defects, domain-wall interactions, or polar nanoregions[33–35]. IR and THz spectroscopy studies connected the loss suppression in RP ferroelectrics to the absence of the central peak, an overdamped excitation around 100 GHz[36]. The current hypothesis relates the suppression of dielectric loss in RP ferroelectrics to mitigation of charged point defects[6,7]. This hypothesis is consistent with the stoichiometry-independent positron lifetime observed in positron annihilation spectroscopy in $(SrTiO_3)_6SrO$[37]. In our survey, the runaway loss at negative bias fields observed in both kHz (**Fig. 3d**) and microwave measurements (**Fig. S24**) indicate an onset of dielectric breakdown associated with migration of charged defects[38]. Furthermore, we can understand the asymmetric tuning curves of the $n = \infty$ perovskite films (**Figs. 4 and S23**) in terms of asymmetric agglomeration of charged defects at the top and bottom electrodes because of their different defect donor chemistry[39] due to their different process history, for which we saw evidence in the EELS measurements (**Fig. S4**).

Nevertheless, relying solely on the interpretation of mitigated charged point defects fails to explain the reversed hierarchy of loss at 10 kHz ($n$ = 8 highest, $n = \infty$ lowest, **Fig. 3d**). Instead, we see evidence for the rock-salt layers *broadening* the dispersion across many frequency decades. In fact, we can clearly discern that the $n$ = 24 film has a broader relaxation than the $n = \infty$ perovskite film (**Fig. S21**). Unfortunately, the dispersions in both the $n$ = 16 and $n$ = 8 films are already too small to reliably confirm this trend. Nonetheless, it stands to reason that the local dielectric response in the RP phase varies with the proximity to both horizontal and vertical rock-salt layers and the resultant local strain. Such local dielectric inhomogeneity could also explain the broadening of the paraelectric-ferroelectric transition with decreasing rock-salt layer periodicity $n$ (**Fig. 3b**). In other words, the sharp relaxation in the perovskite concentrates dielectric loss right at the most relevant application frequencies, whereas the broad relaxation in RPs distributes relatively low loss across many decades of frequencies. Thus, at microwave frequencies, RPs dramatically outperform the perovskite.

There is little reliable tunability and loss data on out-of-plane tunable dielectrics at microwave frequencies. The $Ba_{0.45}Sr_{0.55}TiO_3$ perovskite film ($n = \infty$) serves as a direct internal comparison. The Laue oscillations in the XRD data (**Fig. 2a, inset**) give us confidence that this film has minimal defect concentration in comparison to other $Ba_xSr_{1-x}TiO_3$ films with this barium content. For an external comparison, we benchmark the dielectric tuning figure of merit ($\mathrm{FoM}$) of our films against a 300-nm-thick $Ba_{0.25}Sr_{0.75}TiO_3$ film grown with pulsed-laser deposition[2]. Due to the lower barium content of this film, it has lower loss and smaller tunability than our $n = \infty$ perovskite $Ba_{0.45}Sr_{0.55}TiO_3$. Still, the $Ba_{0.25}Sr_{0.75}TiO_3$ film reaches a relative tunability $T$ = 48 %, though only under an external electric field of 830 kV/cm, which is more than three-times larger than the electric field apply here (250 kV/cm). When compared at the same external electric field (250 kV/cm), the $n$ = 8 film outperforms the $Ba_{0.25}Sr_{0.75}TiO_3$ film by a factor of 10 in the dielectric tuning $\mathrm{FoM}$ at 10 GHz under the same conditions (**Fig. 4d**). Currently, the high resistance of the $SrRuO_3$ bottom electrode limits the overall varactor performance. To make use of the exceptional tunable dielectric performance for achieving exceptional varactor performance, we envision a lift-off process to transfer the RP film to a highly conductive electrode.

## Conclusion

We engineered a MIM-compatible low-loss tunable dielectric by breaking the out-of-plane symmetry of the Ruddlesden-Popper phase. We found that, like their in-plane tunable counterparts, these Ruddlesden-Popper phases suppress dielectric loss at microwave frequencies. First-principles calculations predicted the existence of out-of-plane ferroelectricity in the Ruddlesden-Popper phase of barium strontium titanate $(ATiO_3)_nAO$, $A = Ba_xSr_{1-x}$ for sufficiently dilute rock-salt layers $n$ and high barium content $x$. Phase-field simulations predicted out-of-plane tunability at ambient temperature. We grew Ruddlesden-Popper thin films with $x$ = 0.45 and $n$ = 8, 16, 24, and ∞ and characterized their temperature- and frequency-dependent out-of-plane permittivity, which proved that we can engineer Ruddlesden-Popper thin films to combine out-of-plane tunability with low microwave loss. For the first time, the comparison to a similarly processed perovskite ($n = \infty$) with comparable ferroelectric transition temperature highlights the efficacy of the Ruddlesden-Popper approach to loss suppression. Additionally, the series of rock-salt periodicities demonstrated the critical periodicity to achieve this loss suppression ($16 < n < 24$). A sufficiently low rock-salt periodicity also ensured drastically higher spatial homogeneity of the high frequency dielectric properties across a film. Of the samples studied, the $n = 8$ sample displayed the optimum combination of tunability and loss, recording the highest dielectric tuning figure of merit reported to date for out-of-plane tunable microwave dielectrics at 10 GHz. Engineering out-of-plane tunability into the Ruddlesden-Popper phase overcomes a major obstacle on the path to its industry adoption.

# Methods

## First-principles calculations

All calculations utilize the Vienna Ab-initio Software Package (VASP, version 6.2.0), which implements the projector-augmented wave (PAW) formalism of density functional theory[40–43]. We perform non-spin-polarized calculations using the PBEsol functional and standard PAW potentials from the VASP potential library: Sr_sv ($4s^2$ $4p^6$ $5s^2$), Ba_sv ($5s^2$ $5p^6$ $6s^2$), Ti_sv ($3s^2$ $3p^6$ $3d^2$ $4s^2$) and O ($2s^2$ $2p^4$). Wave functions are expanded up to an energy cutoff of 550 eV, and reciprocal space is sampled on an 8×8×1 Monkhorst-Pack grid (8×8×8 for a perovskite unit cell). We use an electronic convergence criterion of $10^{-8}$ eV.

Conventional Ruddlesden-Popper unit cells with $n$ = 8, 12, 16 and 20 are constructed by extrapolation of the *c*-axis lattice parameter from cells with lower $n$ and insertion of alternating layers of $TiO_2$ and SrO or BaO in accordance with space group symmetry *I*4/*mmm*. Lattice parameters and internal atomic coordinates are allowed to relax until the forces on all atoms are below $5\times10^{-4}$ eV Å$^{-1}$, simulating strain-free thin films. The target compositions of $x$ = 0.3, 0.5, 0.7 are approximated as closely as possible for each value of $n$. Several different cation configurations are sampled, and cation ordering is found to have only a minor effect on the calculated phonon frequencies.

Phonons are calculated at the $\Gamma$ point for the non-polar (*I*4/*mmm*) structures. We used the finite displacement method as implemented in Phonopy (version 2.12.0)[44,45] with VASP as the force calculator. All structures show dynamical instabilities, manifested as imaginary eigenvalues of the dynamical matrix, with the most strongly unstable mode being a polar phonon of $E_{\mathrm{u}}$ symmetry. This mode takes the structure to space group *F*2*mm* with in-plane polarization along [110] referred to the conventional unit cell[46]. For structures with sufficiently high $n$ and $x$ an unstable polar phonon of $A_{2\mathrm{u}}$ symmetry is also found, leading to polarization along [001] and space group *I4mm* if taken alone. The combination of in-plane and out-of-plane polarization leads to space group symmetry *Cm*.

## Phase-field simulations

We employ the dynamical phase-field method to simulate the polarization dynamics and dielectric tunability. The evolution of the polarization is governed by the polarization dynamics equation

$$\mu_{ij}\frac{\partial^2 P_j}{\partial t^2} + \gamma_{ij}\frac{\partial P_j}{\partial t} = -\frac{\delta F}{\delta P_i}, \quad (1)$$

where $\mu_{ij}$ is the polarization effective mass, and $\gamma_{ij}$ is the polarization damping coefficient. We performed the simulations in the low-loss limit, where the damping coefficient ($\gamma_{ij}$ ) has minimal role in the dielectric response. $F$ is the total free energy which is expressed as the integral over volume of the following energy densities:

$$F = \int f_{\mathrm{Landau}} + f_{\mathrm{Elastic}} + f_{\mathrm{Electric}} + f_{\mathrm{Gradient}}\, dV. \quad (2)$$

We elaborate the individual contributions in the **supplementary**. At 1 GHz, damping has minimal impact on the dielectric response, as we perform the simulations in the low-loss limit. We confirmed this limit with frequency-dispersion simulations (**Fig. S2**). The description of dielectric loss of the perovskite may be incomplete in the current phase-field model. Dielectric loss originates from both intrinsic and extrinsic mechanisms. Intrinsic losses arise from anharmonic interactions between phonons and viscous damping, while extrinsic losses arise from the motion of charged defects and the presence of local polar regions[47]. The dynamical phase-field method primarily captures the dynamics of the ferroelectric soft mode and the acoustic modes through the polarization-dynamic and elastodynamic equations[48]. While the current simulations incorporate the intrinsic loss to some extent by accounting for viscous damping and some of the interactions between the soft-mode and acoustic modes, they omit extrinsic contributions, such as defect dynamics and potential local polar regions. Additional loss could transform the predicted under-dampened THz resonance (**Fig. S2**) to the over-dampened GHz relaxation that we observed in the experiment (**Fig. 4a,b**).

Finally, we evaluated the tunability by fitting the permittivity as a function of voltage with a model of field-induced hopping of random, non-interacting dipoles in a double-well potential[49]:

$$\varepsilon(E) = \varepsilon_\infty + \Delta\varepsilon \,\mathrm{sech}(E/E_0)^2 \tag{3}$$

with real parameters $\varepsilon_\infty > 0$, $\Delta\varepsilon > 0$, and $E_0 > 0$. The parameter $E_0$ describes the width of the tuning peak.

## Film growth

We grew the films in a Veeco Gen10* using conventional effusion cells for barium and strontium, an electron-beam evaporator for ruthenium and a Ti-ball* for titanium[50]. XRD and RSM measurements were performed with a PANalytical Empyrean Diffractometer*. We employed a substrate temperature of 650 ˚C for $SrRuO_3$, and 770 ˚C for $(A\mathrm{TiO_3})_nA\mathrm{O}$ (measured by an optical pyrometer operating at a wavelength of 980 nm). We chose 50 nm thick bottom and 25 nm thick $SrRuO_3$ top electrodes which have a similar lattice parameter to our dielectric, because epitaxial electrodes reduce any parasitic interfacial series capacitance. We grew the MIMs on (110)-oriented $DyScO_3$ substrates because its lattice parameter is almost perfectly lattice matched to our active layers: -0.1 % mismatch to $Ba_{0.45}Sr_{0.55}TiO_3$ and +0.6 % mismatch to $SrRuO_3$. To grow these phases on $SrRuO_3$, *in situ* RHEED was essential to accommodate ruthenium deficiency near the surface of the $SrRuO_3$ electrode (for details see **supplementary information**). We deposited the $SrRuO_3$ layers in a background pressure of 1 x $10^{-6}$ Torr of 80 % $O_3$ + 20 % $O_2$ and $(A\mathrm{TiO_3})_nA\mathrm{O}$ layers in a background pressure of 5 x $10^{-7}$ Torr of 10 % $O_3$ + 90 % $O_2$. For $SrRuO_3$, we supplied a ratio of 1 Sr : 3 Ru and utilized the volatility of $RuO_x$ at the substrate temperature (650 °C) to thermodynamically control the stoichiometry of the electrode[51]. After depositing the $SrRuO_3$ back electrode, we cooled the samples to room temperature and stored them in vacuum for ~1 week. We then returned the samples to the growth chamber, heated them to 500 ˚C in 5 x $10^{-7}$ Torr of 10 % $O_3$ + 90 % $O_2$, and supplied (1) 1 monolayer of $TiO_2$ (2) 1 monolayer of SrO, and (3) 1 additional monolayer of $TiO_2$ to the growth surface to encapsulate volatile $RuO_x$ before heating to 770 ˚C to complete the growth of $(ATiO_3)_nAO$.

Our method for growth of $(ATiO_3)_nAO$ has been detailed previously[20,52], but we offer a brief summary here. Between $(AO)_2$ layers, we monitored the √2 x √2 surface reconstruction with *in situ* reflection high-energy electron diffraction (RHEED) to monitor the surface stoichiometry as we alternately supplied monolayers of *A*O and $TiO_2$ such that the surface oscillated between termination with 1 monolayer of *A*O and termination with 2 monolayers of *A*O[20] (see loop in **Fig. S4a**). Each *A*O monolayer was supplied by codepositing flux-matched strontium and barium sources for most of the layer (∼9 s) and briefly closing the barium shutter in the middle of the monolayer (∼2 s) to hit the target stoichiometry (45 % barium). To crystallize a horizontal rock salt layer at this substrate temperature the growth front must be terminated with 3 *A*O layers[23], so we broke the loop in **Fig. S4a** by depositing a 2 monolayer dose of *A*O every $n$ perovskite unit cells[20]. While not reported previously, we used a similar method to enable monitoring of surface stoichiometry in the pure perovskite ($n$ = ∞) sample but never depositing a 2-monolayer dose of *A*O, *i.e.*, never breaking the loop in **Fig. S4a**. This monitored growth method for $Ba_{0.45}Sr_{0.55}TiO_3$ also enables more meaningful comparison because the *A*-site rich growth front is identical to that of RP dielectrics between rock-salt layers. Finally, we cooled the samples to room temperature, stored them in vacuum for ∼1 week, and deposited 25 nm $SrRuO_3$ with the same procedure as the back electrode. The $n$ = 16 and $n$ = 24 samples were processed in parallel, *i.e.*, analogous layers were deposited on the same day, as were the $n$ = 8 and $n$ = ∞ dielectrics. Thus, differences in chamber conditions or calibration accuracy do not drive disparities in loss.

## Scanning Transmission Electron Microscopy

Scanning transmission electron microscopy (STEM) and electron energy loss spectroscopy (EELS) measurements were performed on cross-sectional lamellae prepared via gallium focused ion beam (FIB) lift out on a Thermo Fisher Scientific Helios G4 UX FIB*. High-angle annular dark-field (HAADF) STEM imaging was performed on a FEI Titan Themis* operating at 300 kV with a 21.4 mrad convergence semi-angle, 68 mrad inner collection angle, and 50 pA probe current. Low-angle annular dark-field (LAADF) STEM was performed with a 25 mrad inner collection angle and otherwise identical conditions. Series of rapid-frame images were acquired, aligned and averaged with a rigid registration method optimized to prevent lattice hops to obtain high signal-to-noise ratio, high fidelity atomic resolution images[53]. EELS Elemental maps were recorded with a 965 GIF Quantum ER spectrometer* and a Gatan K2 Summit direct electron detector* operated in counting mode with a 15 pA probe current.

## Device fabrication

We fabricated 4 different device topologies on each of the 4 chips with the RP films (**Fig. S11**). The first device topology is a metal-insulator-metal capacitor for low-frequency measurements. We used the other 3 device topologies for the microwave measurements. These 3 topologies were the surface-short for calibration, the through-short to quantify the bottom electrode contribution, and microwave metal-insulator-metal capacitors to extract the out-of-plane permittivity of the dielectric film[30]. The low-frequency capacitors had a radius of 50 µm. The microwave devices had a signal electrode radius of 15 µm, 20 µm, 25 µm and 30 µm. The inner ground radius was 60 µm and the outer ground radius was 180 µm. To fabricate these devices, we first removed the excess $SrRuO_3$ top electrode with a wet etch (0.1 mol/L solution of $NaIO_4$ in deionized water) to expose the dielectric beneath (**Fig. S12b**). Then, we etched through the dielectric with reactive ion etching to expose the bottom electrode at specific locations (**Fig. S12c**). Next, we deposited nominally 500 nm of gold with conventional lithography techniques to create electrodes that we can contact with on-wafer probes (**Fig. S12 d**).

For calibration purposes, we fabricated a set of coplanar waveguide (CPW) transmission lines on a separate 50.8 mm diameter $(LaAlO_3)_{0.3}(Sr_2TaAlO_6)_{0.7}$ (LSAT) wafer with conventional lithographic techniques and electron-beam vapor deposition. The electrodes were nominally 500 nm thick gold with a 10 nm titanium adhesion layer. The CPWs had a nominally 20 µm wide center conductor, 15 µm wide gap and a 100 µm wide ground plane. The wafer layout included a set of CPW transmission lines with lengths = 0.420 mm, 0.720 mm, 1.040 mm, 2.300 mm, 3.060 mm, 4.000 mm, 5.660 mm, 7.1800 mm, and 9.580 mm, and an offset short-circuit reflect. Additionally, our layout included a series resistor fabricated from a PdAu alloy, which is used for the series resistor calibration and to set the reference impedance on the LSAT substrate.

## Cryogenic Characterization

First, we performed the evaluation of the dielectric tunability at ambient temperature using an impedance analyzer (E4990A, Keysight Technologies*). We contacted the top electrode of the low-frequency MIM capacitors with low-frequency probes and grounded the bottom electrode with a gold pad in the vicinity of the capacitors (**Fig. S13**). We set the AC driving signal to an amplitude of 0.25 kV/cm and set the frequency to 10 kHz frequency while the background DC bias was swept upward from $-V_{\max}$ to $+V_{\max}$. We then repeated the measurement with downward sweeping from $+V_{max}$ to $-V_{max}$ to detect any hysteresis effect. We measured up to $V_{\max}$ of 15 V to 20 V for the RP films and $V_{\max}$ of 2.5 V for the perovskite film. We recorded the capacitance $C$ and calculated the permittivity $\varepsilon_r$ using the parallel-plate-capacitor model

$$\varepsilon_r = \frac{C\,t}{\varepsilon_0\,\pi\,r^2}, \tag{4}$$

where $t$ = 200 nm and $r$ = 25 µm denote the film thickness and the gold electrode radius, respectively. The loss tangent contribution of the $SrRuO_3$ electrode to the device loss is negligible at low frequencies (tan$\delta$ < 0.001). As all capacitors have tan$\delta$ > 0.001, we can identify the device loss as the material loss. We limited the bias electric field on the $n = \infty$ perovskite film to 125 kV/cm (2.5 V) because the loss became so high that we anticipated electrical breakdown. On the $n$ = 24, 16 and 8 RP films, we applied at least 750 kV/cm and up to 1000 kV/cm.

Next, we conducted the temperature-dependent dielectric measurements in a cryogenic probe station (PS-100, Lake Shore Cryotronics*) across the temperature range from 303 K down to 83 K. The amplitude of the AC driving signal was 0.25 kV/cm and we swept the frequency from 100 Hz to 1 MHz. We extrapolated the Curie-Weiss temperature $T_0$ by fitting the high temperature limit of the temperature-dependent permittivity to the Curie-Weiss law

$$\varepsilon_r \sim \frac{1}{T - T_0}. \tag{5}$$

Finally, we performed the *P-E* measurements using a ferroelectric tester (Precision Multiferroic II, Radiant Technologies*) from room temperature down to 30 K. The driving-field frequency was kept at 10 kHz.

## Microwave Characterization

We contacted the high-frequency devices (surface-short, thru-short and microwave MIM capacitors) with 100 µm pitch ground-signal-ground probe tips and measured the scattering parameters (S-parameters) with a vector network analyzer from 40 kHz to 110 GHz (**Fig. S14**). We adopted a two-tier calibration procedure to correct the S-parameters to the plane where the probe tip contacts the DUT. First, we employed a series resistor calibration from 10 MHz to 1 GHz[54] and a multiline-TRL calibration from 1 GHz to 110 GHz[55] as first-tier calibrations with the CPW devices on the LSAT wafer (**Fig. S17**). Then, we employed a surface-short impedance subtraction calibration to correct for the difference between the CPW geometry of the first-tier calibration and the MIM geometry of the test devices[30]. We extracted the permittivity from the corrected S-parameters (**Fig. S18**) with a lumped circuit element model that accounts for the non-homogeneous field distribution in the MIM capacitor that arises due to the high permittivity of the dielectric film and finite conductivity of the bottom electrode[30].

We evaluated the through-shorts to extract the bottom electrode conductivity. Then, we evaluated the capacitors to extract the frequency- and voltage-dependent permittivity. We fitted the permittivity as a function of frequency with the Havriliak-Negami model:

$$\varepsilon(\omega) = \chi_{\mathrm{rlx}} \frac{1}{(1+(i\omega\tau)^{1-\alpha})^{\beta}} + \varepsilon_{f\to\infty}. \tag{6}$$

All fit parameters are real and $\varepsilon_{f\to\infty} > 1$, $\chi_{rlx} > 0$, $\tau > 0$, $0 < \alpha < 1$ and $0 < \beta < 1$. Here, $\chi_{\mathrm{rlx}}$ is dielectric susceptibility of the relaxing mechanism far below its mean relaxation frequency $f_{\mathrm{rlx}} = 1/(2\pi\tau)$. $\varepsilon_{f\to\infty}$ is the permittivity in the high frequency limit far above $f_{\mathrm{rlx}}$. The parameters $\alpha$ and $\beta$ broaden and skew the distribution of relaxation times around the mean relaxation time $\tau$. We only fit the data up to 20 GHz to avoid fitting the increasing noise above 20 GHz.

To evaluate the permittivity as a function of the electric bias field, we fit the permittivity as a function of voltage with a modification of the double-well potential model[49] (**Fig 4c**):

$$\varepsilon(E_{\mathrm{DC}}) = \varepsilon_{\mathrm{V}\to\infty} + \Delta\varepsilon_{\mathrm{tun}} \operatorname{sech}(E_{DC}/E_0)^{2\alpha} + \beta\, E_{\mathrm{DC}}, \tag{7}$$

with real parameters $\varepsilon_{V\to\infty} > 0, \Delta\varepsilon_{\text{tun}} > 0, E_0 > 0$ and additional parameters $\alpha > 0$ and $\beta$. We introduce the dimensionless parameter $\alpha$ for better empirical modeling over a wider electric-field range and the parameter $\beta$ to account for the asymmetry of the tuning curve with respect to negative and positive bias voltages. Some capacitors on the perovskite film ($n = \infty$) become so leaky for $E_{\text{DC}}$ < -100 kV/cm (*i.e.*, when the top electrode was negatively charged) that we did not increase the voltage to avoid dielectric breakdown. Instead, we use the modified double-well potential model (**Eq. 4**) to extract the asymmetry parameter $\beta$ and the relative tunability at $E_{\max}$ = +250 kV/cm bias field via:

$$T_{\max} = 1 - \frac{\varepsilon_r(E_{\max})}{\varepsilon_r(0\ \text{kV/cm})}. \tag{8}$$

To extract the tuning peak width $E_0$, we set $\alpha$ = 1 and only fit within $|E|$ < 75 kV/cm because $\alpha$ and $E_0$ are strongly correlated fitting parameters that both describe the broadness of the tuning peak. We calculate the material quality factor of the films at $E$ = 0 kV/cm via:

$$Q_0 = 1/\tan\delta(0\ \text{kV/cm}). \tag{9}$$

We assess the performance of tunable materials with the commonly used dielectric tuning figure of merit (**Fig 4d**),

$$\text{FoM} = Q_0 T_{\max}, \tag{10}$$

even though we are aware of other metrics for comparison[56]. In this work, the focus is on the material performance in contrast to the overall device performance. This focus implies that the FoM does not include the electrode loss and we consider the tunability at the same external electric field rather than comparing the overall device loss and the tunability at the same voltage.

Finally, we chose representative microwave MIM capacitors for **Fig. 4** from the ones in **Fig. S20**. We made this choice based on the agreement of the fit with the data. For some $n$ = 8 capacitors, the fit resulted in a material quality factor $Q_0$ > 200 at 10 GHz (**Fig. S25**). In fact, this is below the resolution of our measurement method, which we estimate to be at $Q_0$ = 100 without fitting the dispersion model and $Q_0$ = 200 with fitting.

## Figures, tables, captions

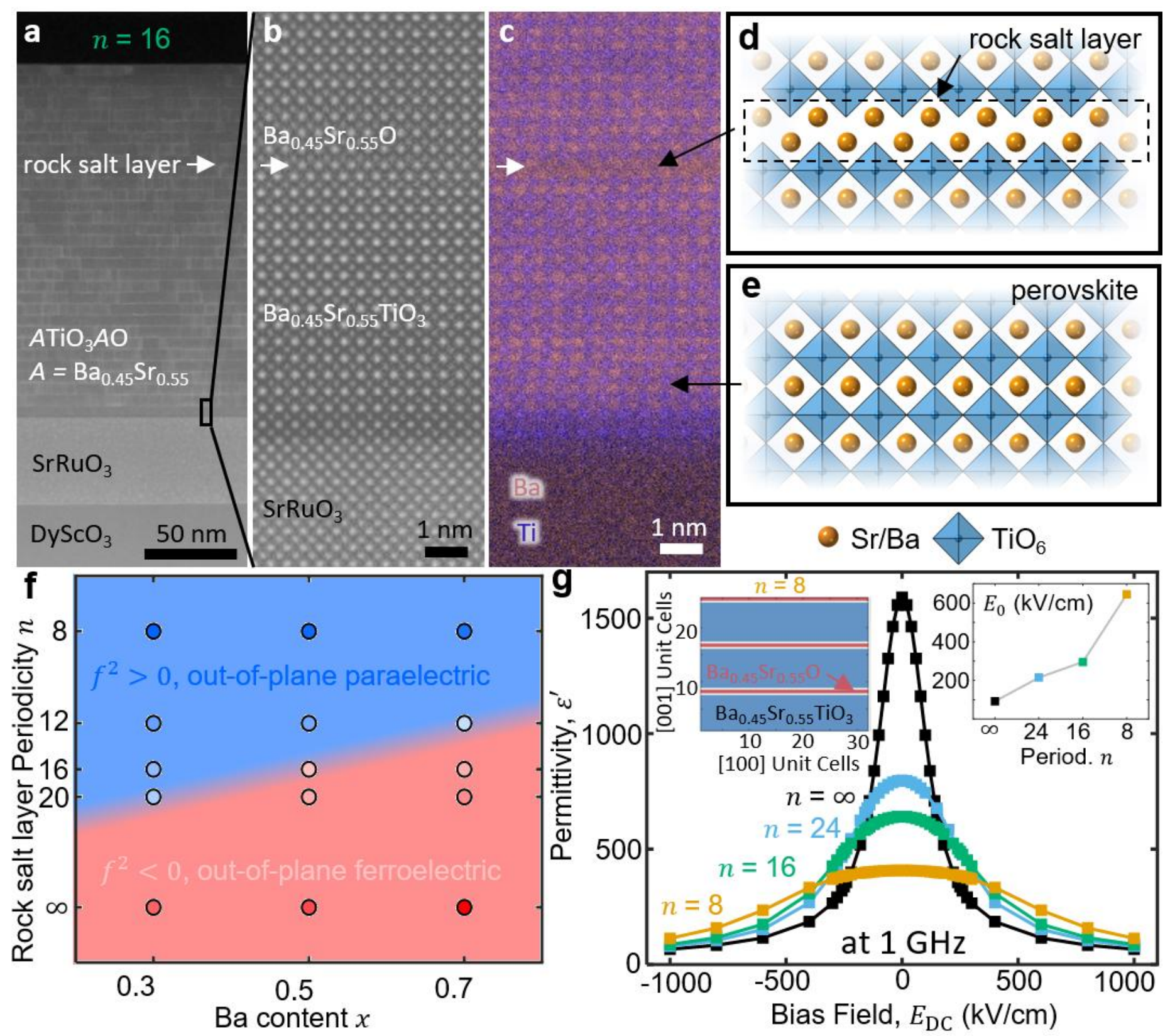

**Figure 1 | The RP phase and predictions about its dielectric properties.** **a**, Low-magnification cross-sectional LAADF-STEM image, **b**, atomic-resolution cross-sectional HAADF-STEM image and **c**, EELS elemental map of the heterostructure in cross-section containing $(A\mathrm{TiO_3})_{16}A\mathrm{O}$, $A$ = $\mathrm{Ba_{0.45}Sr_{0.55}}$. The lamella for the STEM images comes from a region with an etched top $\mathrm{SrRuO_3}$ electrode. **d**,**e** Diagrams of a rock salt layer and the perovskite, respectively. **f**, First-principles prediction of out-of-plane ferroelectricity for high barium content $x$ and large periodicities $n$. Blue and red correspond to a positive and negative squared phonon frequency, respectively. **g**, Phase-field simulation predicting tunability at ambient temperature with insets showing the simulation setup (left) and the tuning peak width $E_0$ as a function of the periodicity $n$ (right).

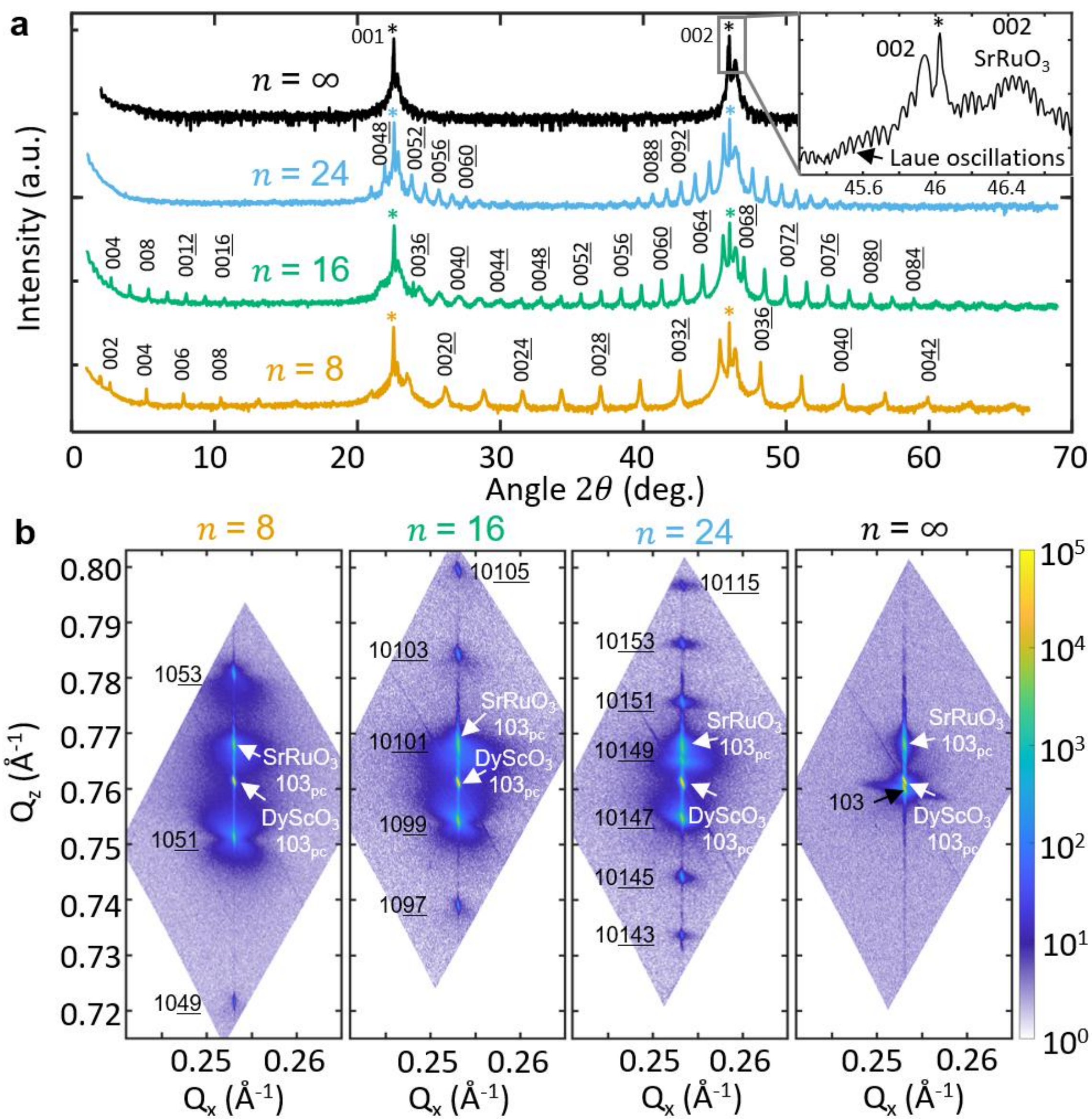

**Figure 2 | X-ray diffraction characterization.** **a**, XRD $\theta$-$2\theta$ scans of the heterostructures containing $(ATiO_3)_nAO$ dielectrics. Underlined numbers correspond to a single index, and asterisks (*) indicate peaks from the substrate. The inset shows the high-resolution XRD $\theta$-$2\theta$ scan near the 002 peak of the $n = \infty$ heterostructure. **b**, Reciprocal space maps near the 103 pseudo-cubic peaks of each heterostructure. Black indices indicate peaks arising from the dielectric rock-salt layer.

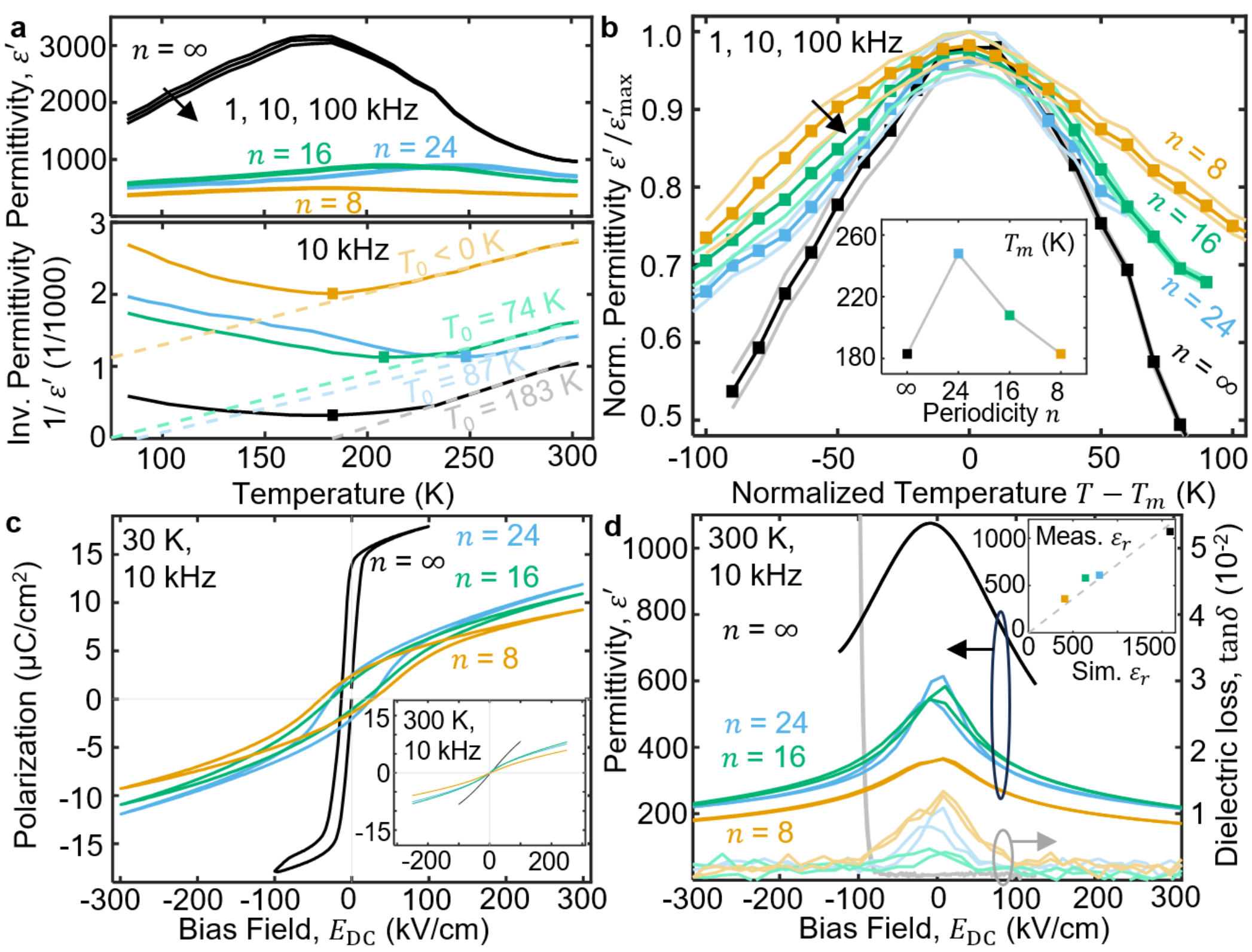

**Figure 3 | Cryogenic characterization**. **a**, Temperature-dependent real permittivity (top) of the four RP $x$ = 0.45 films ($n$ = 8, 16, 24, and ∞) and the inverse of the real permittivity (bottom) for extraction of the Curie-Weiss temperature $T_0$. Points represent the temperature of maximum permittivity $T_\mathrm{m}$. **b**, Normalized permittivity as a function of the temperature difference to $T_\mathrm{m}$, showing the slight frequency dispersion and the broadening of the phase transition with increasing periodicity $n$. The inset shows the nonmonotonic trend of $T_\mathrm{m}$ as a function of periodicity $n$. **c**, Polarization - electric field loops at 30 K with inset at 300 K, revealing the ferroelectric and paraelectric behavior, respectively. **d**, Tunability of the complex permittivity (real permittivity on left axis, loss tangent on right axis) for the four RP films at ambient temperature (300 K) with inset comparing the measured permittivity to the value predicted by phase-field simulations.

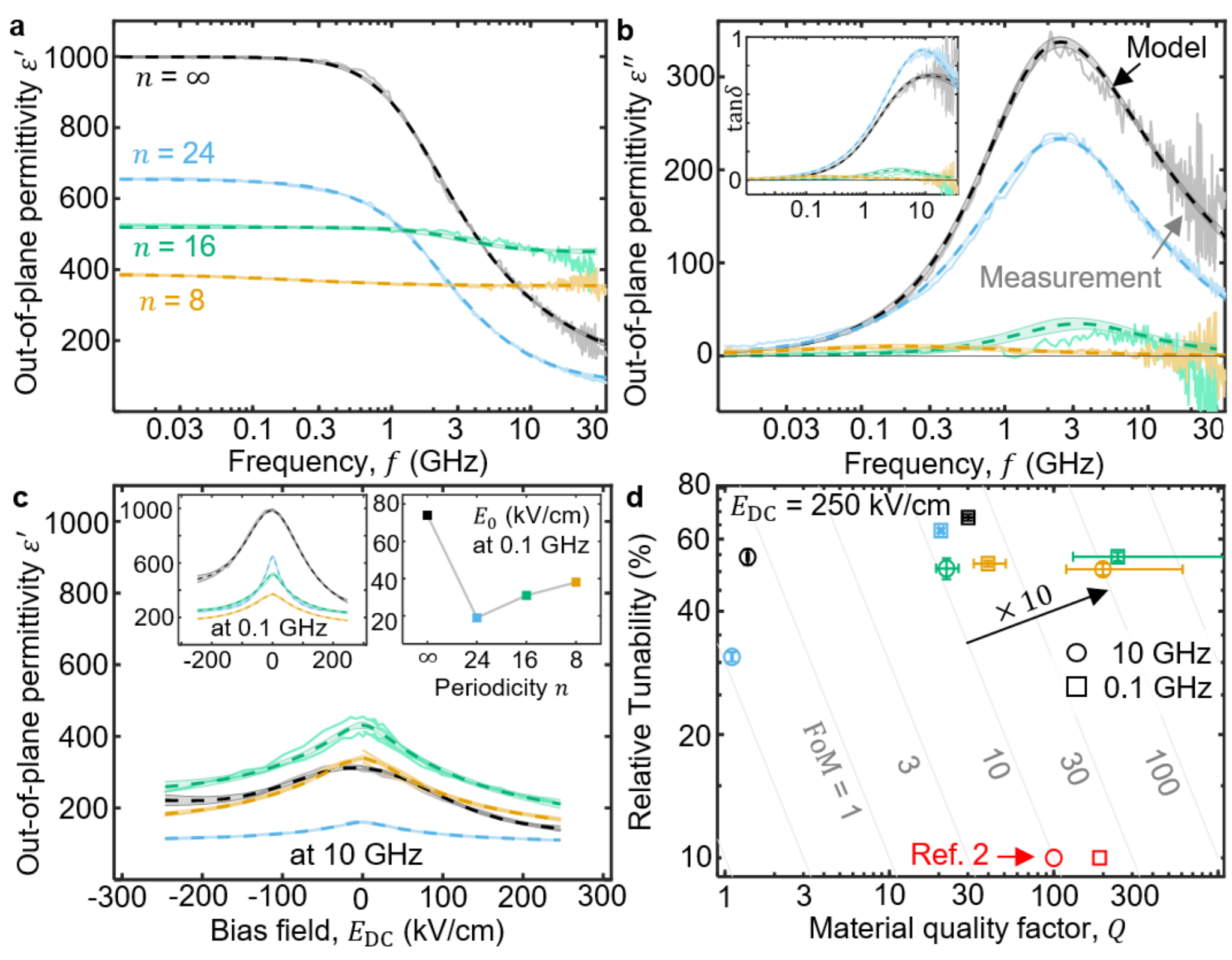

**Figure 4 | Microwave characterization at ambient temperature**. Frequency dispersion of the **a**, real part of the permittivity ($\varepsilon'$) and **b**, imaginary part of the relative permittivity ($\varepsilon''$, with inset showing the loss tangent $\tan\delta$) of selected microwave MIM capacitors of the four RP $x$ = 0.45 films ($n$ = 8, 16, 24, and $\infty$). The dashed lines are Havriliak-Negami fits, the shaded areas show their standard uncertainty. **c**, Tunability of the microwave permittivity at 10 GHz with insets showing the tunability at 0.1 GHz (left) and the fitted tuning peak width $E_0$ (right). **d**, Material performance map comparing the material quality factor $Q_0$ at 0 V (horizontal axis), relative tunability $T_{\max}$ at 5 V (vertical axis) and the figure of merit $\mathrm{FoM} = Q_0 T_{\max}$ (tilted isolines) at 0.1 GHz and 10 GHz. Extracted from selected microwave MIM capacitors in comparison to a previously reported high performance out-of-plane tunable microwave dielectric[2] (we show a map of all MIM capacitors in **Fig. S25**).

## Acknowledgement

The authors thank the critical review of A. Hagerstrom, C. J. Long, J. Jargon, and J. C. Booth, all with the National Institute of Standards and Technology (NIST), and Y. A. Birkhölzer at Cornell University for their critical feedback during this research, and their comments on this manuscript, and A. Osella (also with NIST) for support in the clean room. This paper is an official contribution of NIST, not subject to copyright in the US. The figure data are available at the NIST Public Data Repository (https://doi.org/10.18434/mds2-3983), additional data are available from the corresponding author upon reasonable request.

* Certain commercial equipment, instruments, software, or materials, commercial or non-commercial, are identified in this paper in order to specify the experimental procedure adequately. Such identification does not imply recommendation or endorsement of any product or service by NIST, nor does it imply that the materials or equipment identified are necessarily the best available for the purpose.

G.H.O. acknowledges Betül Pamuk at Williams College for inspiring discussions and Craig J. Fennie at Cornell University for foundational work on ferroelectric Ruddlesden-Popper phases and for providing computational resources for the DFT calculations. M.R.B., D.S., L.W.M. and D.G.S. acknowledge that this research was sponsored by the Army Research Laboratory and was accomplished under Cooperative Agreement Number W911NF-24-2-0100. The views and conclusions contained in this document are those of the authors and should not be interpreted as representing the official policies, either expressed or implied, of the Army Research Laboratory or the U.S. Government. The U.S. Government is authorized to reproduce and distribute reprints for Government purposes notwithstanding any copyright notation herein. Z.T., L.-Q.C., and L.W.M. acknowledges the support of the U.S. Department of Energy, Office of Science, Office of Basic Energy Sciences, under Award Number DE-SC-0012375. A.S. acknowledges the National Science Foundation under Grant DMR-2329111. This research was conducted with support from the Cornell University Center for Advanced Computing. A.R. acknowledges the support of the National Science Foundation Graduate Research Fellowship Program under Grant No. DGE1255832. S.K. acknowledges support by the Czech Science Foundation (Project No. 24-10791S) and by the project TERAFIT (Project No. CZ.02.01.01/00/22_008/00045910004594), co-financed by the European Union and the Czech Ministry of Education, Youth and Sports. V.G. acknowledges support from the Martina Roeselová Memorial Fellowship. D.T. acknowledges support in part by National Science Foundation Grant DMR-2104918. This work made use of the synthesis and electron microscopy facilities of the Platform for the Accelerated Realization, Analysis, and Discovery of Interface Materials (PARADIM), which are supported by the National Science Foundation under Cooperative Agreement No. DMR-2039380, and the Cornell Center for Materials Research shared instrumentation facility. The FEI Titan Themis 300 was acquired through NSF-MRI-1429155, with additional support from Cornell University, the Weill Institute and the Kavli Institute at Cornell.

## Supplementary Information

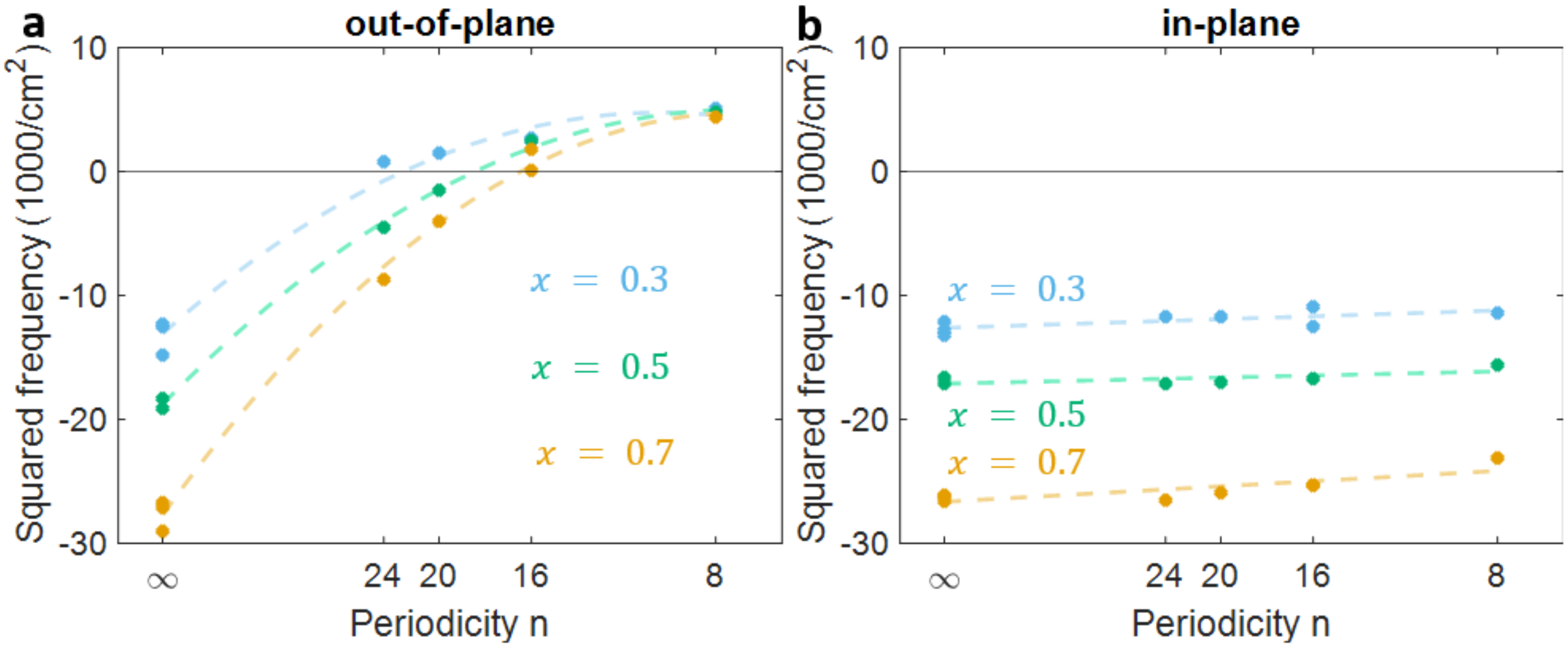

**Figure S1 |** Squared phonon frequency as a function of the rock salt layer periodicity *n* from Density Functional Theory calculations for **a**, out-of-plane phonons ($A_{2u}$ symmetry) and **b**, in-plane phonons ($E_u$). The different squared frequencies at one periodicity $n$ correspond to different cation orderings. The lines are guides to the eye only.

### Phase-field simulations

Following the Landau-Ginzburg-Devonshire Theory of Ferroelectricity[57], the Landau free-energy density describes the intrinsic stability of the ferroelectric phase with respect to the paraelectric phase and is written as a Taylor expansion of the polarization about the paraelectric phase (*i.e.*, $P_i$ = 0):

$$f_{\mathrm{Landau}} = a_{11}(T)(P_1^2 + P_2^2 + P_3^2) + a_{1111}(P_1^4 + P_2^4 + P_3^4) + a_{1122}(P_1^2 P_2^2 + P_2^2 P_3^2 + P_1^2 P_3^2)$$

$$\ldots + a_{111111}(P_1^6 + P_2^6 + P_3^6) + a_{111122}\left(P_1^4(P_2^2 + P_3^2) + P_2^4(P_1^2 + P_3^2) + P_3^4(P_1^2 + P_2^2)\right)$$

$$... + a_{112233}P_1^2P_2^2P_3^2 + a_{11111111}(P_1^8 + P_2^8 + P_3^8)$$

$$... + a_{11111122}\left(P_1^6(P_2^2 + P_3^2) + P_2^6(P_1^2 + P_3^2) + P_3^6(P_1^2 + P_1^2)\right)$$

$$\ldots + a_{11112222}(P_1^4P_2^4 + P_2^4P_3^4 + P_1^4P_3^4) + a_{11112233}(P_1^4P_2^2P_3^2 + P_2^4P_1^2P_3^2 + P_3^4P_1^2P_2^2),$$

where $a_{ij}$, $a_{ijkl}$, $a_{ijklmn}$ and $a_{ijklmnop}$ are the dielectric stiffness coefficients under constant stress conditions. The gradient energy density is

$$f_{\text{Gradient}} = \frac{1}{2}G_{ijkl}\frac{\partial P_i}{\partial x_j}\frac{\partial P_k}{\partial x_l},$$

where $G_{ijkl}$ is the gradient energy tensor. The elastic energy density is

$$f_{\text{Elastic}} = \frac{1}{2}c_{ijkl}\left(\varepsilon_{ij} - \varepsilon_{ij}^0\right)\left(\varepsilon_{kl} - \varepsilon_{kl}^0\right),$$

where $c_{ijkl}$ is the elastic stiffness tensor, $\varepsilon_{ij}$ is the total strain (using the stress-free paraelectric phase as a reference) and $\varepsilon_{ij}^0$ is the eigenstrain which is related to the polarization by

$$\varepsilon_{ij}^0 = Q_{ijkl}P_kP_l,$$

where $Q_{ijkl}$ is the electrostrictive tensor. The elastic stiffness and electrostrictive coefficients are given in **Table S3**. For simplicity, we assume the elastic stiffness coefficients are homogenous throughout the superlattice. To calculate the local strain, we use the mechanical displacement ($u_i$) where the evolution of the mechanical displacement is given by the electrodynamic equation

$$\rho\frac{\partial^2 u_i}{\partial t^2} = \nabla \cdot \left(\sigma_{ij} + \beta\frac{\partial \sigma_{ij}}{\partial t}\right),$$

where $\rho$ is the material mass density, $\beta$ is the stiffness damping coefficient and $\sigma_{ij}$ is the stress field.

The electrostatic energy density is

$$f_{\text{Electric}} = -E_i P_i - \frac{1}{2}\epsilon_0 \kappa_{ij}^b E_i E_j,$$

where $\epsilon_0$ is the vacuum permittivity and $\kappa_{ij}^b$ is the background dielectric constant which is chosen as 20 and is isotropic. More details on numerical methods for the dynamical phase field method can be found in the literature[48]. All the coefficients are given in **Table S3**.

We define the RP structure by spatially varying the material properties (*i.e.*, dielectric stiffness coefficients, etc.) For numerical stability, the interface between the rock salt and perovskite layer is smoothed using a hyperbolic tangent function. The interface between the ferroelectric film and $DyScO_3$ (110) surface is treated as coherent allowing for the lattice mismatch strain to be calculated using the equivalent cubic lattice constant ($a_{BSTO}^c = 3.955$ Å) and the pseudocubic lattice parameters of the substrate $0.5\sqrt{a^2+b^2} = 3.952$ Å, $0.5\,c = 3.946$ Å[58]. Since the lattices of the film and substrate are orthogonal, there is no in-plane misfit shear strain such that:

$$\varepsilon_{11} = \frac{a_{DSO}^{[110]_O} - a^c}{a^c}, \varepsilon_{22}\frac{a_{DSO}^{[001]_O} - a^c}{a^c}, \varepsilon_{12} = \varepsilon_{21} = 0.$$

A system size of $64\Delta x_1 \times 2\,\Delta x_2 \times 3(2n+1)\Delta x_3$, where $n$ is the number of perovskite layers in the RP unit cell. For $n$ = ∞, a system size of $64\Delta x_1 \times 2\,\Delta x_2 \times 51\Delta x_3$ is used. Periodic boundary conditions were applied in all directions. In all simulations $\Delta x_1 = \Delta x_2 = \Delta x_3 = 0.2$ nm. Random noise is used to initiate the simulation with a magnitude of 0.01 C/m$^2$. To simulate the dielectric tunability, a combination of a static bias field and a 1 GHz sinusoidal electric field with an amplitude of 0.01 kV/cm is applied to the system. The real and imaginary dielectric constant is found by fitting the time-dependent polarization response to a sine and cosine function.

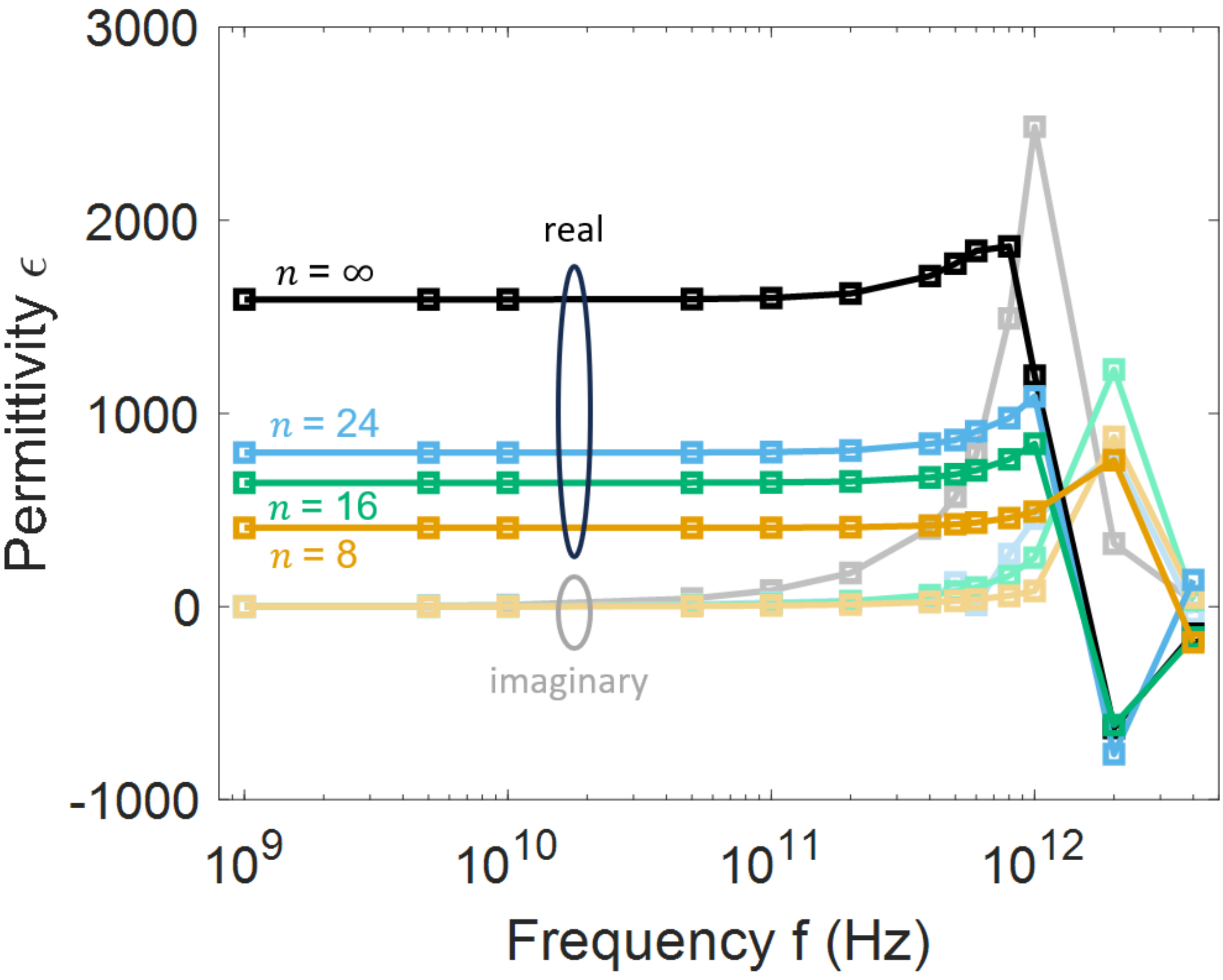

**Figure S2 |** Frequency dispersion of the phase field simulation for the 4 RP $A\mathrm{TiO_3}A\mathrm{O}$, $A$ = $\mathrm{Ba_{0.45}Sr_{0.55}}$ films ($n$ = 8, 16, 24, and ∞).

**Table S3 |** Phase-field simulation parameters.

| Coefficient | $Ba_{0.45}Sr_{0.55}TiO_3$ | $Ba_{0.45}Sr_{0.55}O$ |
|---|---|---|
| $a_{11}$ | $5.234 \times 10^{7}\left(\frac{\mathrm{J}}{\mathrm{m}^3}\ \frac{\mathrm{m}^4}{\mathrm{C}^2}\right)$ | $2.98 \times 10^{9}\left(\frac{\mathrm{J}}{\mathrm{m}^3}\ \frac{\mathrm{m}^4}{\mathrm{C}^2}\right)$ |
| $a_{1111}$ | $2.097 \times 10^{7}\left(\frac{\mathrm{J}}{\mathrm{m}^3}\ \frac{\mathrm{m}^8}{\mathrm{C}^4}\right)$ | $0\left(\frac{\mathrm{J}}{\mathrm{m}^3}\ \frac{\mathrm{m}^8}{\mathrm{C}^4}\right)$ |
| $a_{1122}$ | $4.674 \times 10^{8}\left(\frac{\mathrm{J}}{\mathrm{m}^3}\ \frac{\mathrm{m}^8}{\mathrm{C}^4}\right)$ | $0\left(\frac{\mathrm{J}}{\mathrm{m}^3}\ \frac{\mathrm{m}^8}{\mathrm{C}^4}\right)$ |
| $a_{111111}$ | $1.294 \times 10^{9}\left(\frac{\mathrm{J}}{\mathrm{m}^3}\ \frac{\mathrm{m}^{12}}{\mathrm{C}^6}\right)$ | $0\left(\frac{\mathrm{J}}{\mathrm{m}^3}\ \frac{\mathrm{m}^{12}}{\mathrm{C}^6}\right)$ |
| $a_{111122}$ | $-1.75 \times 10^{9}\left(\frac{\mathrm{J}}{\mathrm{m}^3}\ \frac{\mathrm{m}^{12}}{\mathrm{C}^6}\right)$ | $0\left(\frac{\mathrm{J}}{\mathrm{m}^3}\ \frac{\mathrm{m}^{12}}{\mathrm{C}^6}\right)$ |
| $a_{112233}$ | $-3.10 \times 10^{9}\left(\frac{\mathrm{J}}{\mathrm{m}^3}\ \frac{\mathrm{m}^{12}}{\mathrm{C}^6}\right)$ | $0\left(\frac{\mathrm{J}}{\mathrm{m}^3}\ \frac{\mathrm{m}^{12}}{\mathrm{C}^6}\right)$ |
| $a_{11111111}$ | $3.863 \times 10^{10}\left(\frac{\mathrm{J}}{\mathrm{m}^3}\ \frac{\mathrm{m}^{16}}{\mathrm{C}^8}\right)$ | $0\left(\frac{\mathrm{J}}{\mathrm{m}^3}\ \frac{\mathrm{m}^{16}}{\mathrm{C}^8}\right)$ |
| $a_{11111122}$ | $2.529 \times 10^{10}\left(\frac{\mathrm{J}}{\mathrm{m}^3}\ \frac{\mathrm{m}^{16}}{\mathrm{C}^8}\right)$ | $0\left(\frac{\mathrm{J}}{\mathrm{m}^3}\ \frac{\mathrm{m}^{16}}{\mathrm{C}^8}\right)$ |
| $a_{11112222}$ | $1.637 \times 10^{10}\left(\frac{\mathrm{J}}{\mathrm{m}^3}\ \frac{\mathrm{m}^{16}}{\mathrm{C}^8}\right)$ | $0\left(\frac{\mathrm{J}}{\mathrm{m}^3}\ \frac{\mathrm{m}^{16}}{\mathrm{C}^8}\right)$ |
| $a_{11112233}$ | $1.367 \times 10^{10}\left(\frac{\mathrm{J}}{\mathrm{m}^3}\ \frac{\mathrm{m}^{16}}{\mathrm{C}^8}\right)$ | $0\left(\frac{\mathrm{J}}{\mathrm{m}^3}\ \frac{\mathrm{m}^{16}}{\mathrm{C}^8}\right)$ |
| $c_{11}$ | $2.53 \times 10^{11}$ (Pa) | $2.53 \times 10^{11}$ (Pa) |
| $c_{12}$ | $0.98 \times 10^{11}$(Pa) | $0.98 \times 10^{11}$(Pa) |
| $c_{44}$ | $1.20 \times 10^{11}$(Pa) | $1.20 \times 10^{11}$(Pa) |
| $Q_{11}$ | 0.0813 ($\mathrm{m}^4/\mathrm{C}^2$) | 0.0 ($\mathrm{m}^4/\mathrm{C}^2$) |
| $Q_{12}$ | −0.0224 ($\mathrm{m}^4/\mathrm{C}^2$) | 0.0 ($\mathrm{m}^4/\mathrm{C}^2$) |
| $Q_{44}$ | 0.0183 ($\mathrm{m}^4/\mathrm{C}^2$) | 0.0 ($\mathrm{m}^4/\mathrm{C}^2$) |
| $G_{11}$ | $1.04 \times 10^{-10}$ ($\mathrm{J\,m}^3/\mathrm{C}^2$) | $1.04 \times 10^{-10}$ ($\mathrm{J\,m}^3/\mathrm{C}^2$) |
| $G_{12}$ | 0 ($\mathrm{J\,m}^3/\mathrm{C}^2$) | 0 ($\mathrm{J\,m}^3/\mathrm{C}^2$) |
| $G_{44}$ | $0.52 \times 10^{-10}$ ($\mathrm{J\,m}^3/\mathrm{C}^2$) | $0.52 \times 10^{-10}$ ($\mathrm{J\,m}^3/\mathrm{C}^2$) |
| $\rho$ | $6.02 \times 10^{3}$($\mathrm{kg/m}^3$) | $6.02 \times 10^{3}$($\mathrm{kg/m}^3$) |
| $\beta$ | $6.0 \times 10^{-12}$(s) | $6.0 \times 10^{-12}$(s) |
| $\kappa^b$ | 10 (unitless) | 10 (unitless) |
| $\mu$ | $1.35 \times 10^{-18}\left(\frac{\mathrm{kg}}{\mathrm{m}}\frac{\mathrm{m}^4}{\mathrm{C}^2}\right)$ | $1.35 \times 10^{-18}\left(\frac{\mathrm{kg}}{\mathrm{m}}\frac{\mathrm{m}^4}{\mathrm{C}^2}\right)$ |
| $\gamma$ | $5.88 \times 10^{-6}\left(\frac{\mathrm{kg}}{\mathrm{ms}}\frac{\mathrm{m}^4}{\mathrm{C}^2}\right)$ | $5.88 \times 10^{-6}\left(\frac{\mathrm{kg}}{\mathrm{ms}}\frac{\mathrm{m}^4}{\mathrm{C}^2}\right)$ |

## Film growth and structural characterization

Our precise awareness of the surface stoichiometry proved vital to the first successful synthesis of these $(A\mathrm{TiO_3})_nA\mathrm{O}$ atop $SrRuO_3$. When we initiated growth of the dielectrics at 770 °C, the unexpected persistence of the √2 x √2 RHEED reconstruction indicated that the surfaces were Ti deficient, requiring manual compensation of approximately ½ monolayer of $TiO_2$. We interpret that, despite our attempted encapsulation, $RuO_x$ near the electrode's surface was depleted, and Ti was displaced to fill Ru vacancies. The integrated EELS intensity near the interface between $(A\mathrm{TiO})_{16}A\mathrm{O}$ and the $SrRuO_3$ back electrode (**Fig. S4b**) indicates that titanium penetrated ½ unit cell deeper into the electrode than $A$ = $Ba_{0.45}Sr_{0.55}$, which supports this hypothesis. Because we monitored the surface stoichiometry with *in situ* RHEED, we were able to precisely compensate for $RuO_x$ evaporation to prevent nucleation of the thermodynamically stable $Ba_2TiO_4$ phase[18], which occurs when the surface becomes too $A$-site rich. After a few manually controlled perovskite unit cells, we ran the recipe deduced from calibration and observed the expected evolution of the √2 x √2 RHEED reconstruction[20] (**Fig. S4a**). Periodically, we checked the √2 x √2 RHEED reconstruction and made small adjustments to shutter times (~1 %) to accommodate drifting fluxes.

Interestingly, we observed hourglass-shaped diffuse scattering in the RSMs, which monotonically decreases with the rock-salt periodicity $n$ (**Fig. 2b**). However, we do not attempt more rigorous quantitative analysis of the diffused scattering because the data was taken with a line-focused lab source, which can add artifacts that would undermine the analysis.

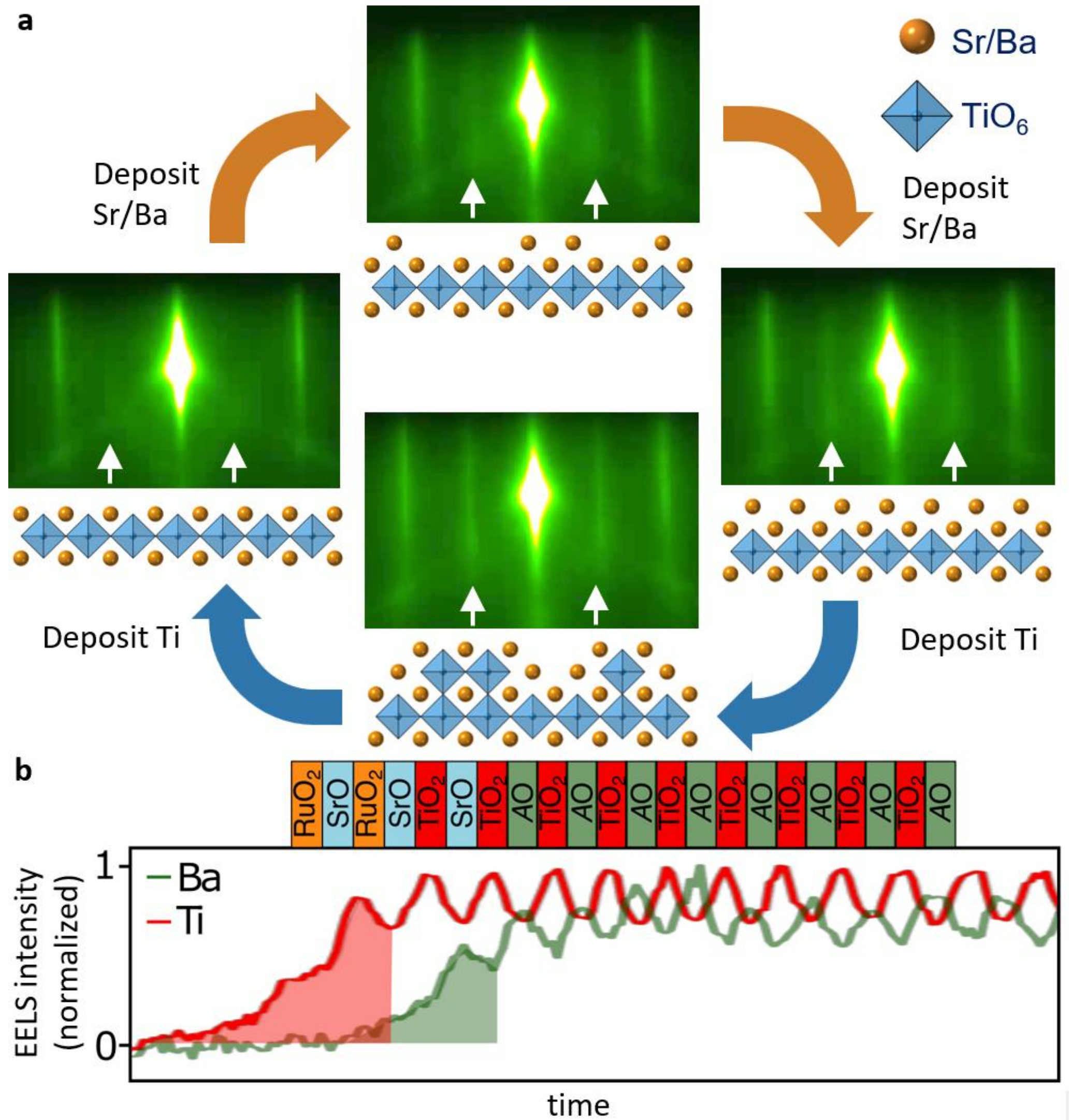

**Figure S4 | a**, Diagram of the growth process at different stages, with a RHEED image and surface sketch at each stage. The arrows in the RHEED image point at the monitored half-order peak indicative of a √2 x √2 surface reconstruction. **b**, Normalized EELS intensity of Ba and Ti near the interface between $(A\mathrm{TiO})_{16}A\mathrm{O}$ and the $SrRuO_3$ back electrode (corresponding to EELS image in **Fig. 1c** from the main text). The diagram above indicates the intended layer sequence (including the encapsulating $SrTiO_3$ unit cell), and the shaded portion indicates the integrated EELS intensity of Ba (green) and Ti (red) beyond the intended depth.

## In-plane Infrared and THz spectroscopy

We measured IR reflectivity spectra with a FTIR Bruker IFS 113v spectrometer in the frequency range of 30 $cm^{-1}$ to 3300 $cm^{-1}$ (*i.e.*, 1 THz to 100 THz). Additionally, we obtained THz dielectric spectra with a custom-made time-domain THz transmission spectrometer in the range of 10 $cm^{-1}$ to 50 $cm^{-1}$ (*i.e.*, 0.3 THz to 1.5 THz). The IR reflectivity and time domain THz transmission spectra allowed us to detect polar phonons and their dielectric contributions to permittivity[36]. Both the IR and the THz measurements are only sensitive to the in-plane permittivity of the sample. Because of the substrate anisotropy, we measured the IR spectra with a polarization of $\boldsymbol{E} \parallel [001]$ and $\boldsymbol{E} \parallel [1\bar{1}0]$, with respect to the crystal axes of the (110) oriented $DyScO_3$ substrate. We characterized two 200 nm thick RP $(ATiO_3)_nAO$, $A = Ba_{0.45}Sr_{0.55}$ thin films with $n$ = 8 and $n$ = 24 (**Fig. S6**). The $DyScO_3$ substrate (thickness 398 μm) was opaque for the THz radiation with the polarization $\boldsymbol{E} \parallel [001]$. Therefore, we only obtained the THz spectra for the polarization $\boldsymbol{E} \parallel [1\bar{1}0]$, while we measured the IR reflectivity spectra for both polarizations.

The THz spectra revealed a strong contribution of the central mode (*i.e.*, overdamped excitations below phonons) near the phase transition temperature. The softening (i.e., red shift) of both excitations towards 175 K causes a dielectric anomaly in the static permittivity $\varepsilon'(0)$ (*i.e.*, the permittivity given by the sum of the contributions of all polar excitations) (**Fig. S7**). This trend indicates an in-plane ferroelectric phase transition at 175 K in both the $n$ = 8 and the $n$ = 24 films.

Since we see a static permittivity anomaly in the plane of the sample, while other low-frequency and microwave measurements have revealed a dielectric out-of-plane anomaly, we can argue that the ferroelectric phase has a polarization oblique to the plane of the films. This is consistent with the results of first-principles calculations, which predict a monoclinic *Cm* structure in the films.

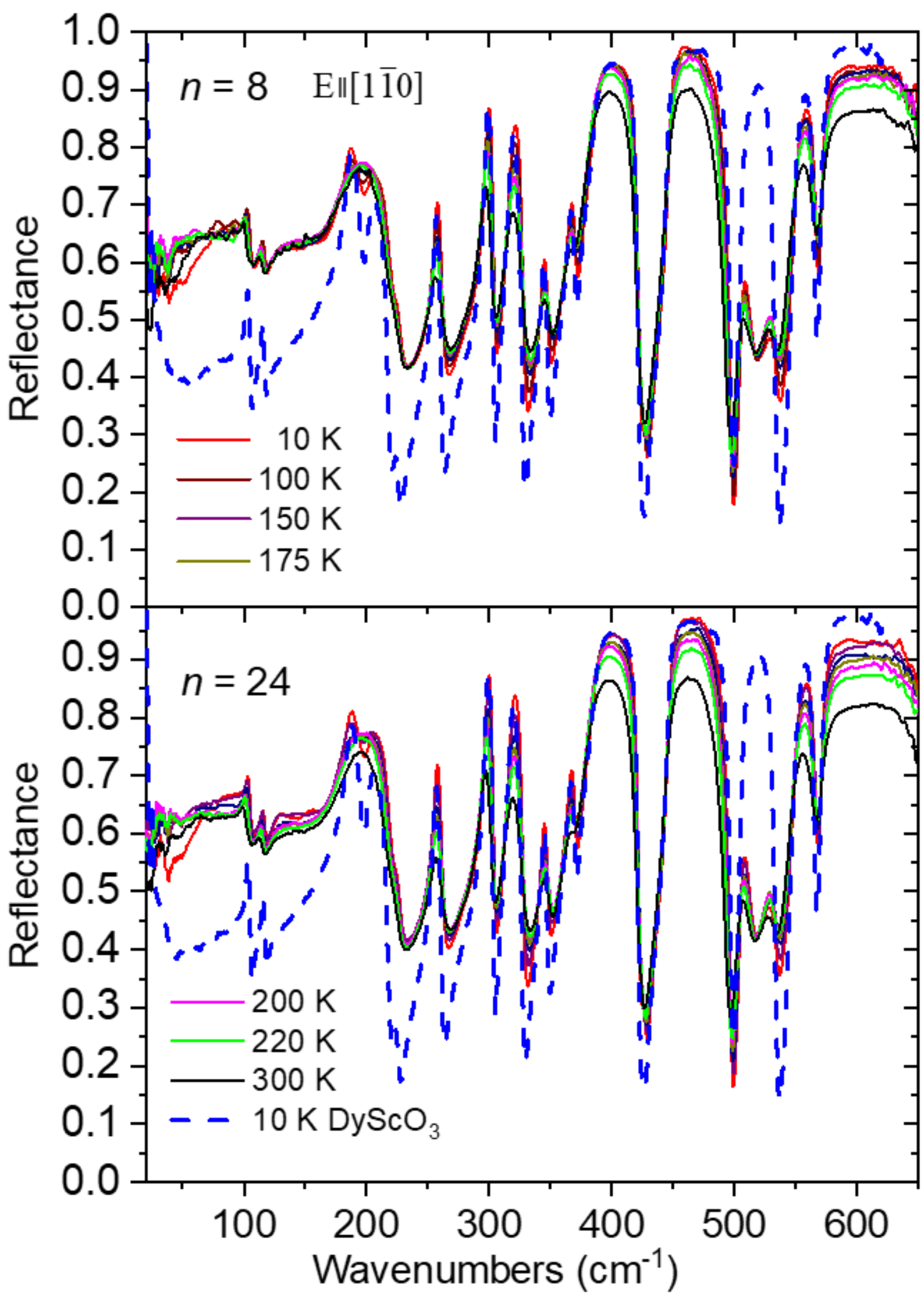

**Figure S5 |** Temperature dependent $\boldsymbol{E} \parallel [1\bar{1}0]$ polarized IR reflectance spectra of 200 nm thick $n$ = 8 and $n$ = 24 thin films. The dashed blue line represents the 10 K $\boldsymbol{E} \parallel [1\bar{1}0]$ polarized IR spectrum of $DyScO_3$ substrate for comparison. The direction is that of orthorhombic $DyScO_3$ substrate. The different reflectance intensities of the bare substrate and film grown on the substrate are due to the phonon contributions from the film.

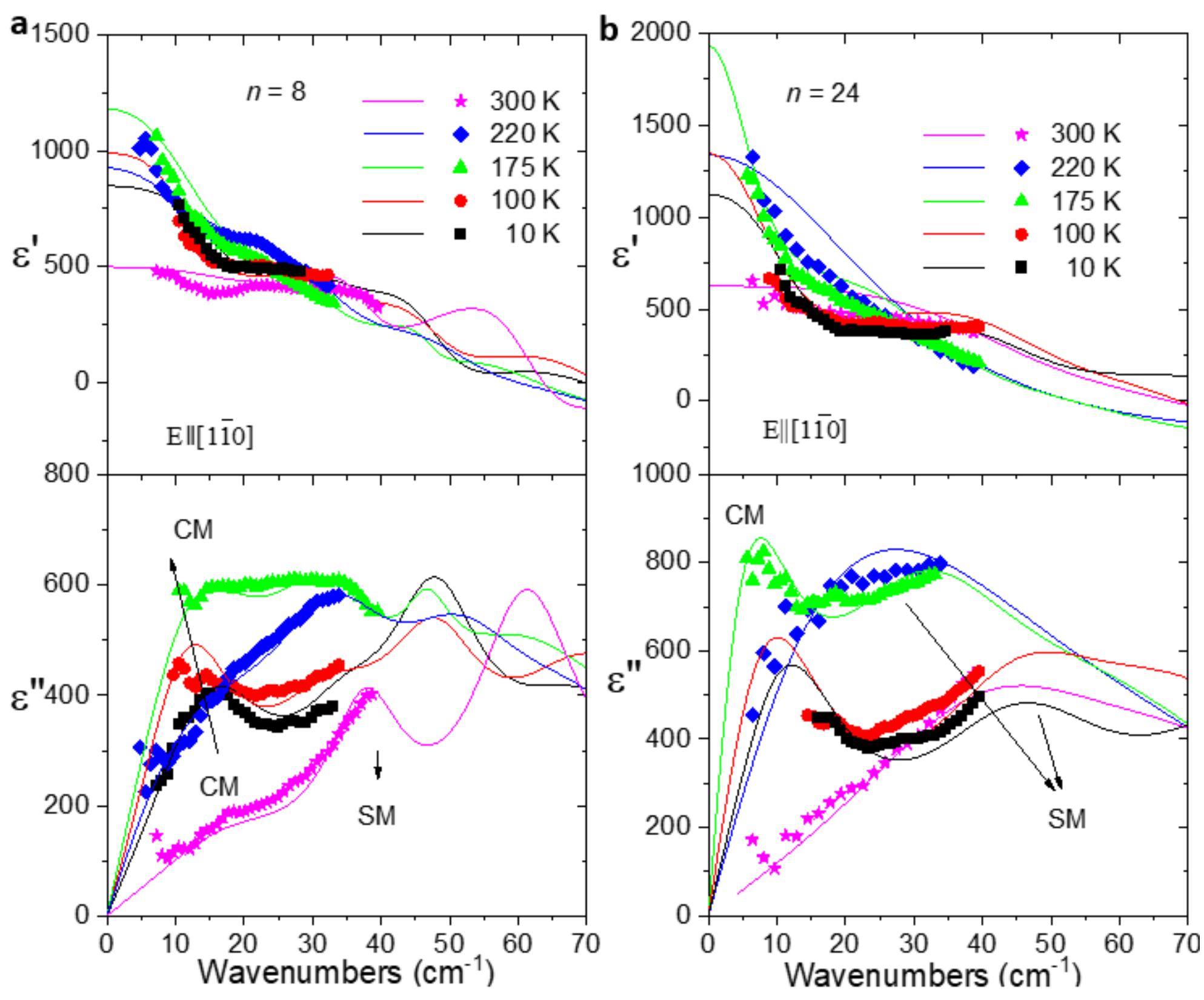

**Figure S6 |** Complex in-plane permittivity spectra of 200 nm thick $n$ = 8 **a**, and $n$ = 24 **b**, thin films obtained from $\boldsymbol{E} \parallel [1\bar{1}0]$ polarized THz (symbols) and low-frequency part of IR spectra (lines). The direction is that of orthorhombic $DyScO_3$ substrate. CM and SM label the central mode and the soft mode, respectively.

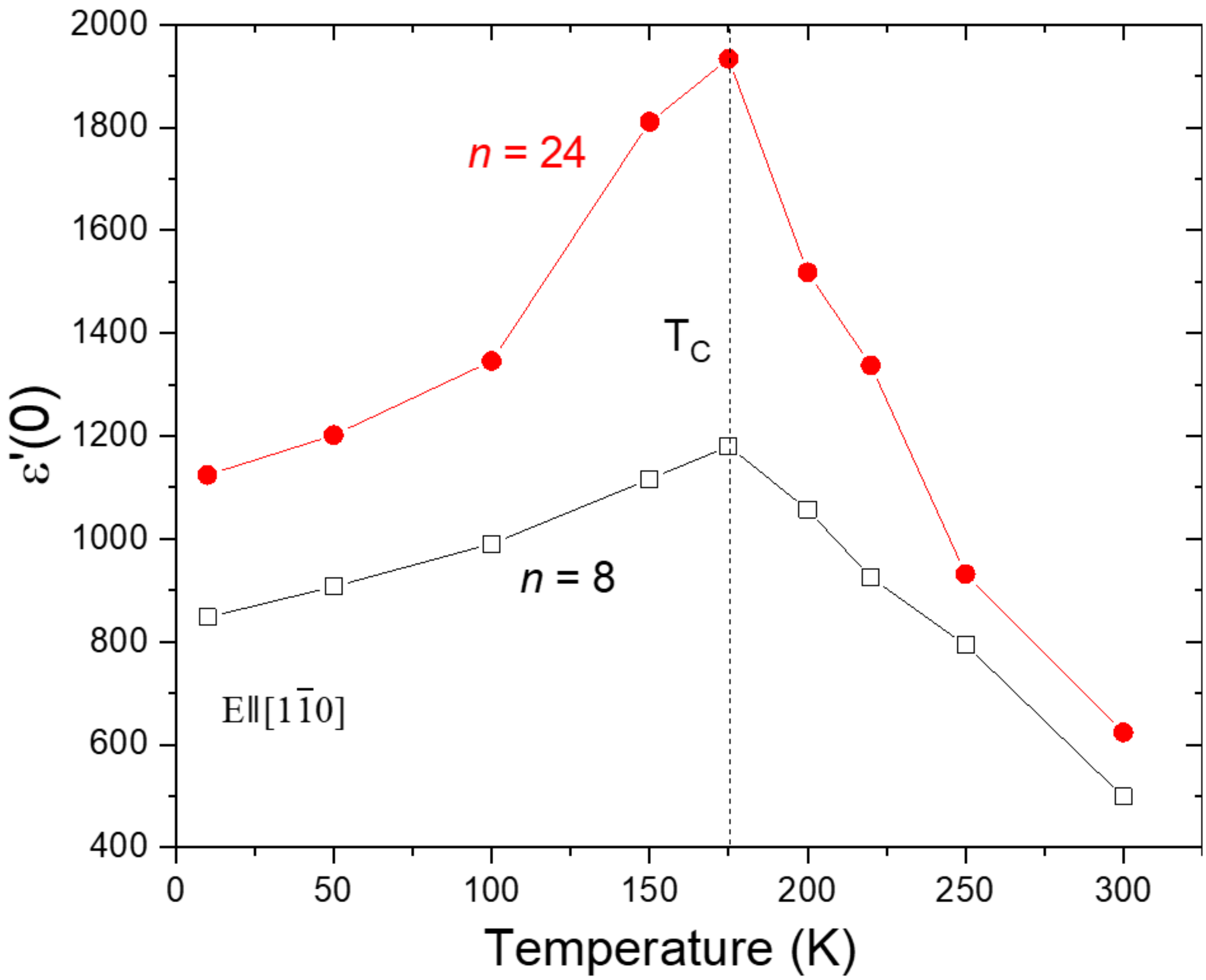

**Figure S7 |** Temperature dependence of the static permittivity for $\boldsymbol{E} \parallel [1\bar{1}0]$ polarization in both $n$ = 8 and $n$ = 24 thin films obtained from the sum of dielectric contributions of all polar phonons and central mode. The vertical line marks the probable transition temperature $T_c$ = 175 K.

## UV Raman spectroscopy

We applied variable-temperature ultraviolet Raman spectroscopy to study barium strontium titanate RP thin film superlattices $(A\mathrm{TiO_3})_n A\mathrm{O}$, $A$ = $\mathrm{Ba_{0.45}Sr_{0.55}}$ with variable $n$ grown by molecular-beam epitaxy on $\mathrm{SrRuO_3/DyScO_3}$ substrates. We measured the spectra in backscattering geometry normal to the film surface using a Jobin Yvon T64000* triple spectrometer equipped with a liquid nitrogen cooled multichannel charge coupled device detector. We used an ultraviolet excitation (325 nm line of He-Cd laser) for excitation to reduce the substrate contribution. We recorded spectra in the temperature range of 20 K to 350 K using a variable temperature closed cycle helium cryostat. The maximum laser power density was less than 0.5 $\mathrm{W/mm^2}$ at the sample surface, low enough to avoid any noticeable local heating of the sample.

An ideal perovskite structure like bulk $\mathrm{SrTiO_3}$ is centrosymmetric, and all phonon modes have odd symmetry with respect to the inversion center. Therefore, these phonons are not active in the first-order Raman spectra since they are prohibited by the selection rules, and the spectrum of bulk cubic perovskites such as $\mathrm{SrTiO_3}$ only contains the second order (two-phonon) features. However, a breakdown in the inversion symmetry of the crystal (*e.g.* due to a ferroelectric distortion), leads to the appearance of the first-order phonon peaks in the Raman spectra.

All samples show the presence of polar phonon modes indicating that the films are ferroelectric at low temperatures, with out-of-plane polarization (**Fig. S8**). The spectra of the $n = \infty$ perovskite film show peaks close to those observed previously in $\mathrm{Ba_{0.5}Sr_{0.5}TiO_3}$ crystals[59]. From the temperature evolutions of the Raman spectra, we determined the ferroelectric phase transition $T_\mathrm{c}$ as the temperature for which the phonon peak intensity drops to near zero (**Fig. S9**). We mostly used the $\mathrm{TO_4}$ peak near 535 $\mathrm{cm^{-1}}$ to determine $T_\mathrm{c}$, as it indicates the phase transition most clearly. (The shoulder of the $\mathrm{LO_3}$ peak shows the same decrease with temperature). We determined $T_c$ through plotting the integrated Raman intensity normalized by the corresponding intensity at 20 K and the Bose-Einstein temperature factor. The value of the $T_c$ is determined by the linear fit of the data near the transition. Some residual Raman activity persists above the $T_c$ due to deviation of perfect symmetry selection rules, it does not show characteristic linear decrease and is not related to the ferroelectric polarization. All films studied had $T_c$ temperatures below room temperature. The $T_c$ varied between $\sim$ 190 K and 200 K. We estimate the accuracy of the $T_c$ determination from Raman data to be ±25 K.

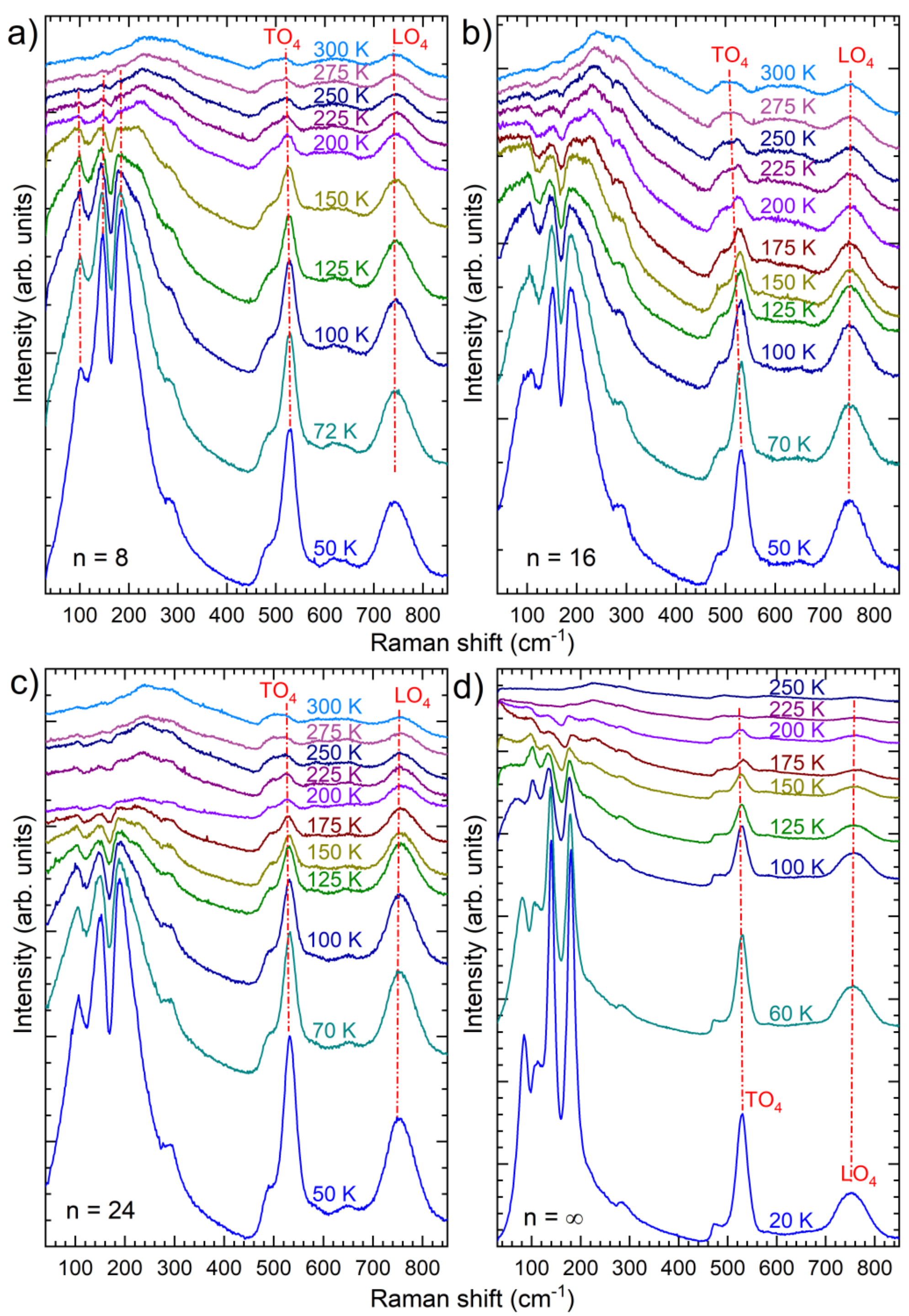

**Figure S8 |** Temperature evolution of the Raman spectra for four RP films $(ATiO_3)_nAO$, $A = Ba_{0.45}Sr_{0.55}$ ($n$ = 8, 16, 24, and ∞). Red dashed lines are guides for the eye.

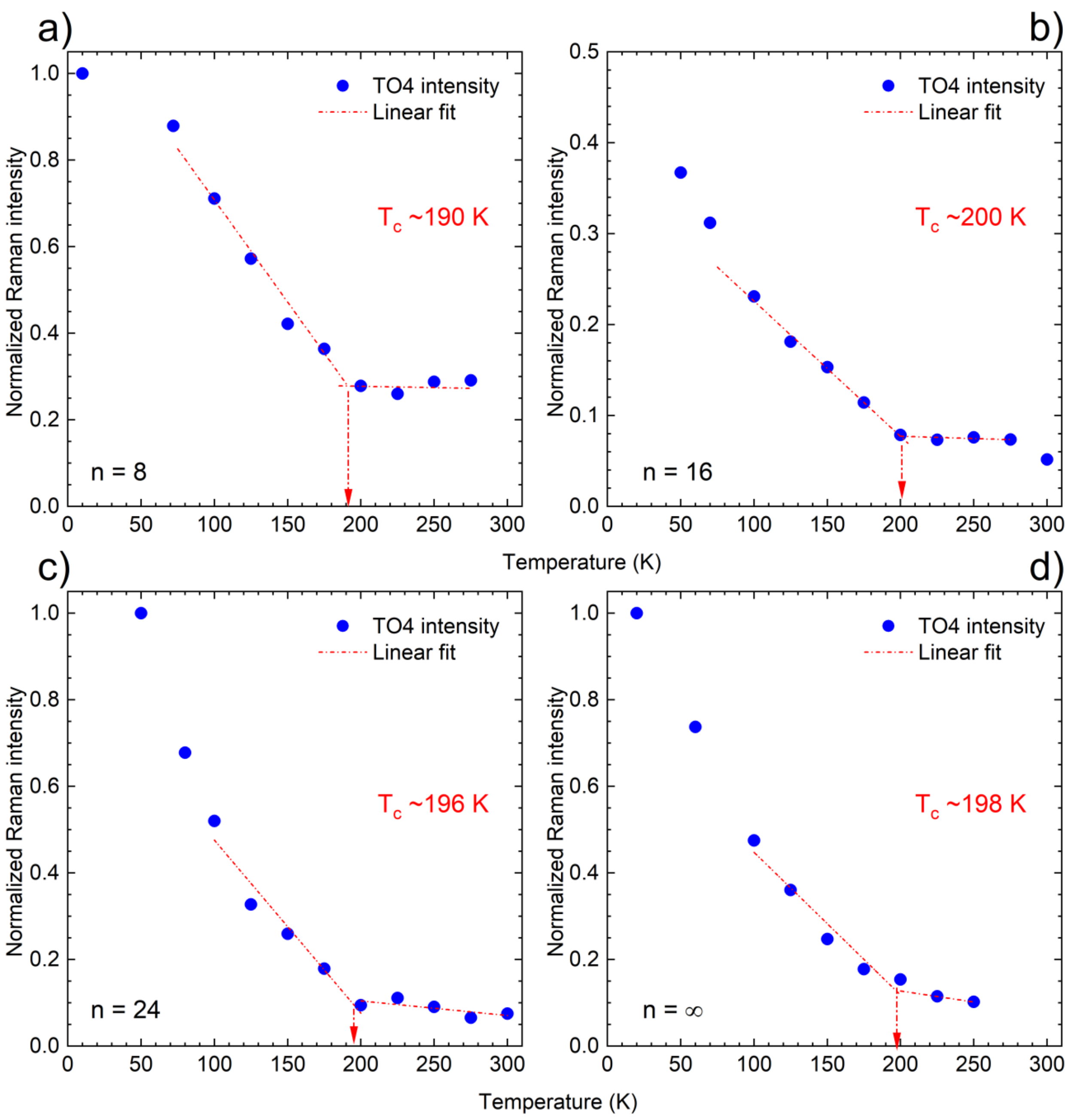

**Figure S9 |** Temperature dependence of the normalized Raman intensity for the $TO_4$ peak of the 4 RP films $(ATiO_3)_nAO$, $A = Ba_{0.45}Sr_{0.55}$ ($n$ = 8, 16, 24, and ∞).

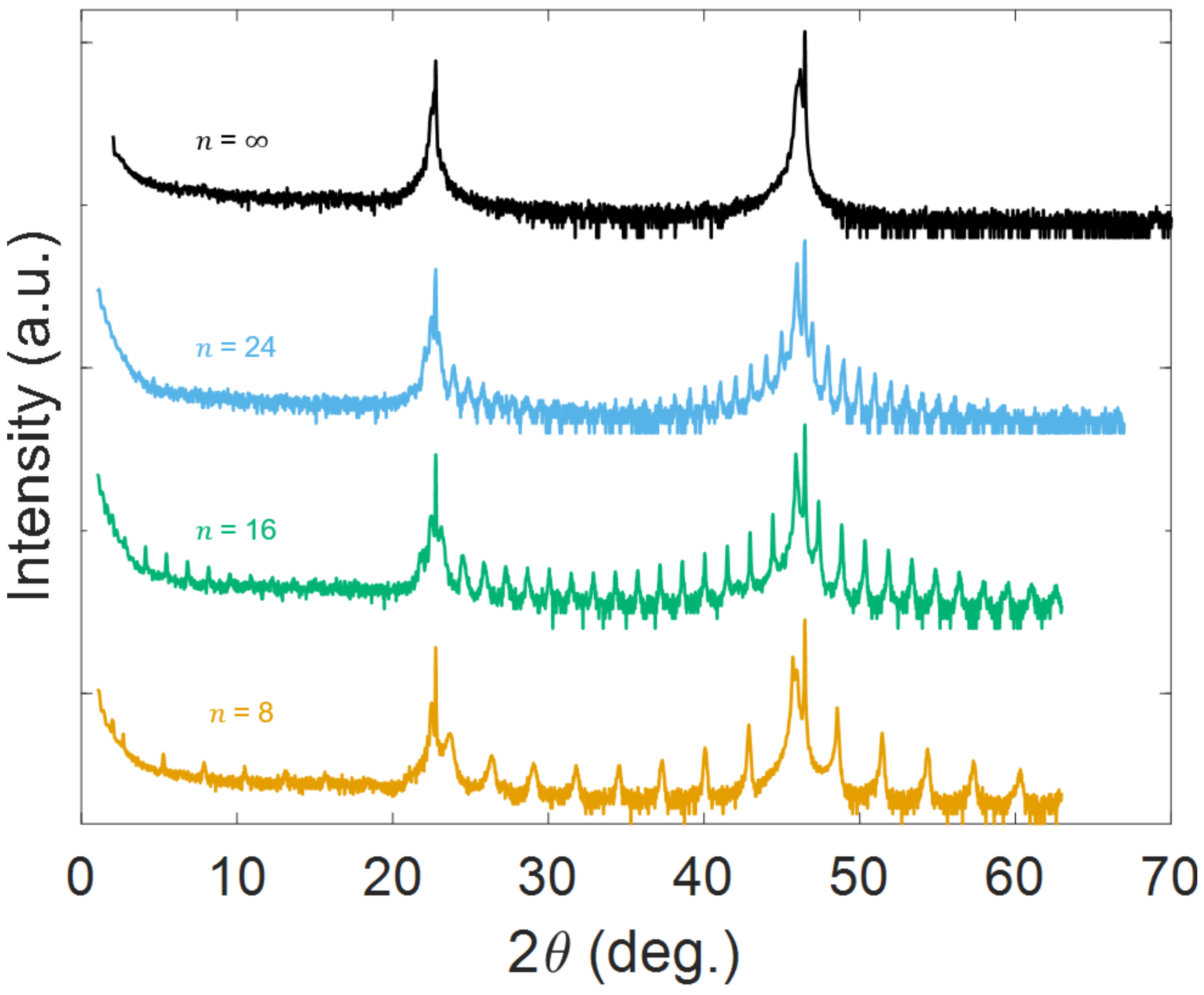

**Figure S10 |** XRD of four RP $(Ba_{0.15}Sr_{0.85}TiO_3)_nSrO$ films ($n$ = 8, 16, 24, and ∞).

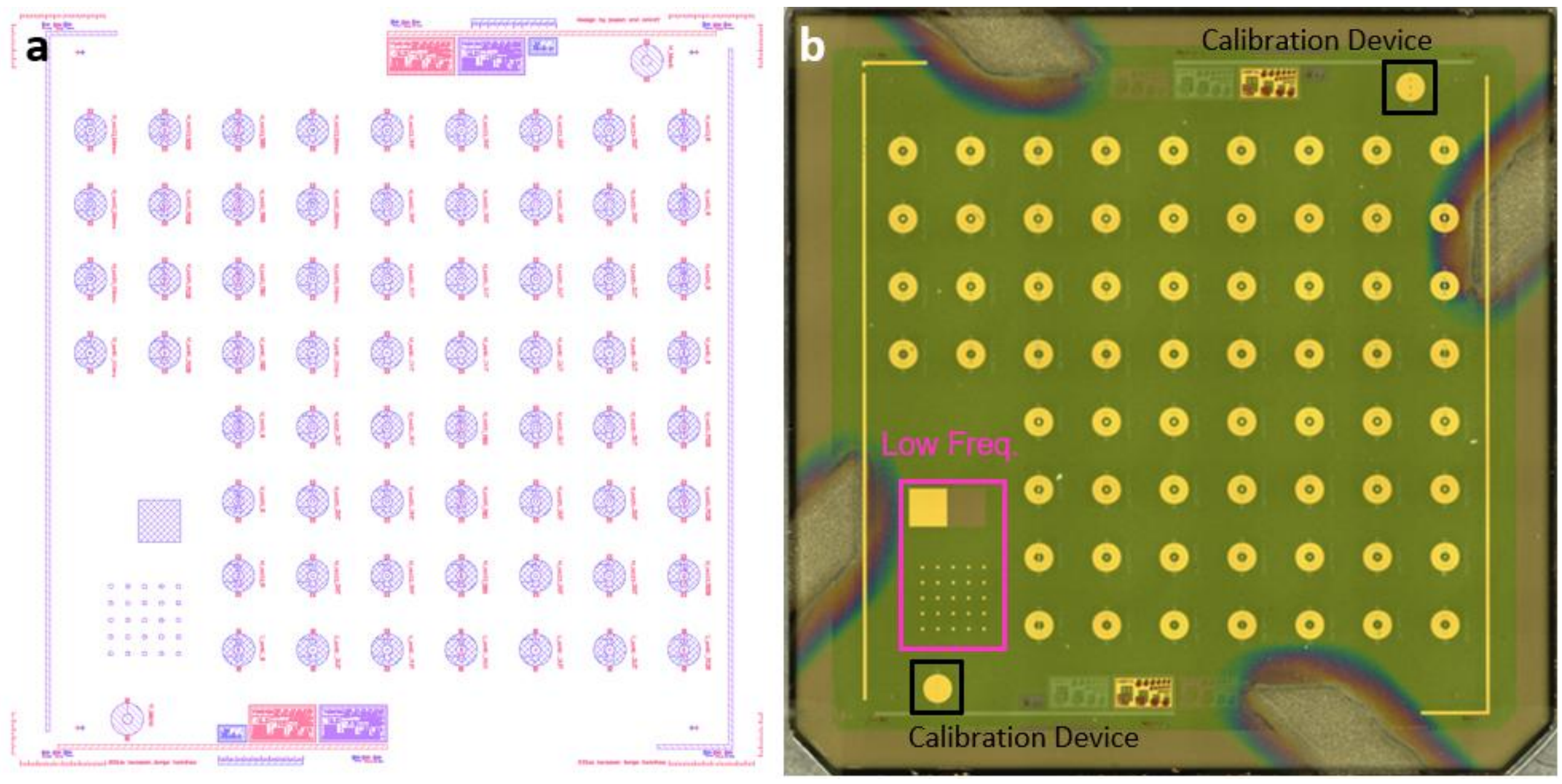

**Figure S11 |** Layout file **a** and image of fabricated chip **b**. The boxes highlight the low-frequency capacitors and the microwave-frequency calibration devices.

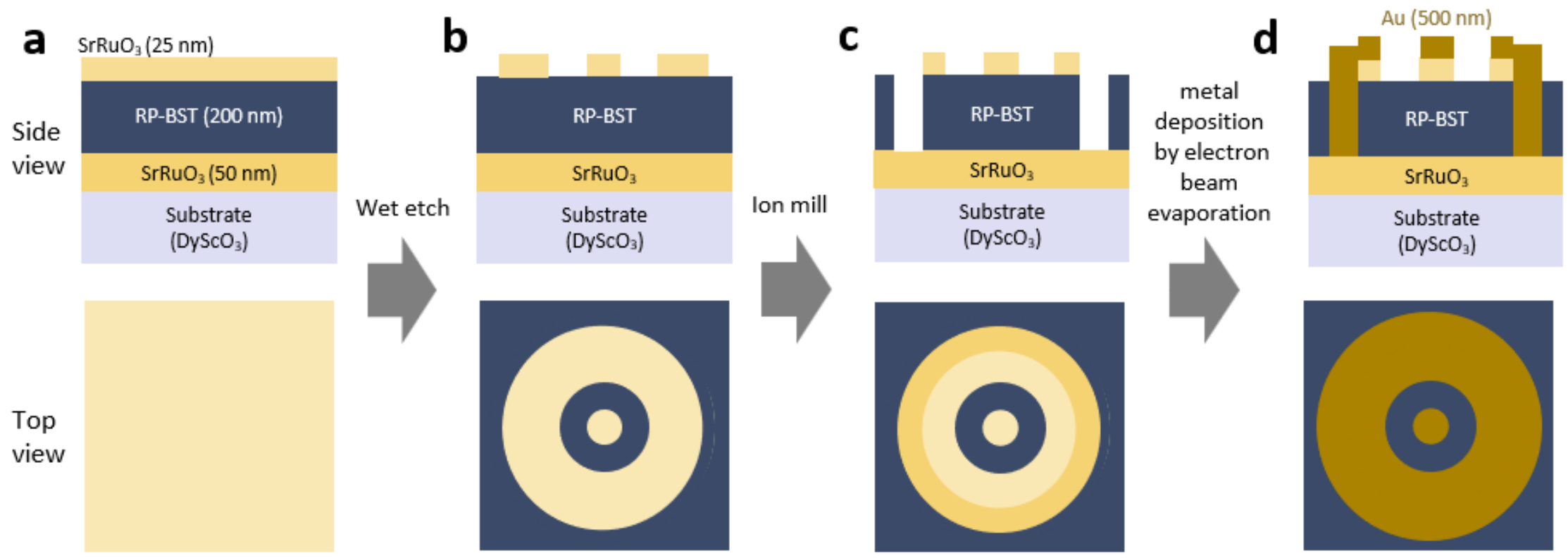

**Figure S12 |** Schematic of the device fabrication process. **a**, Pristine sample after film growth, **b**, after wet etch, **c**, after ion mill and **d**, after gold deposition (final device).

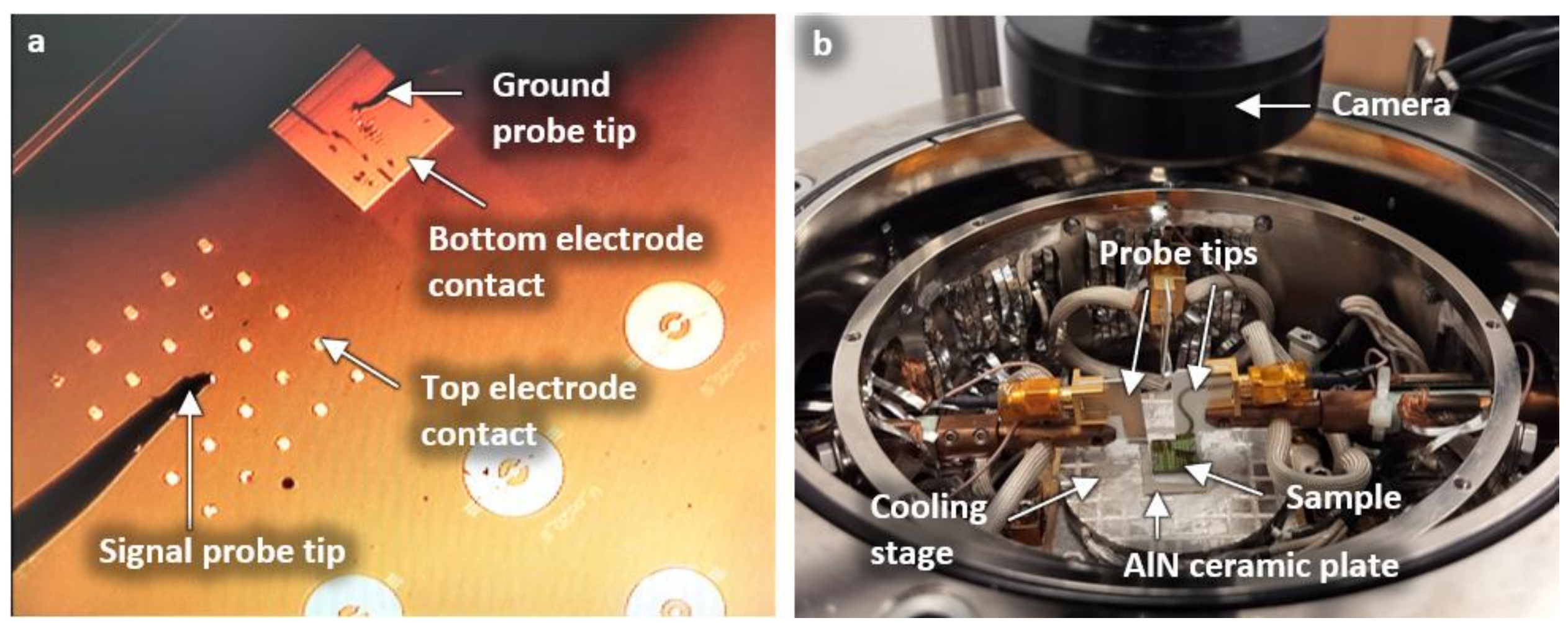

**Figure S13 |** Low-frequency cryogenic measurement setup. **a**, Microscope image of the landed probes on a device. **b**, Chip mounted in probe station.

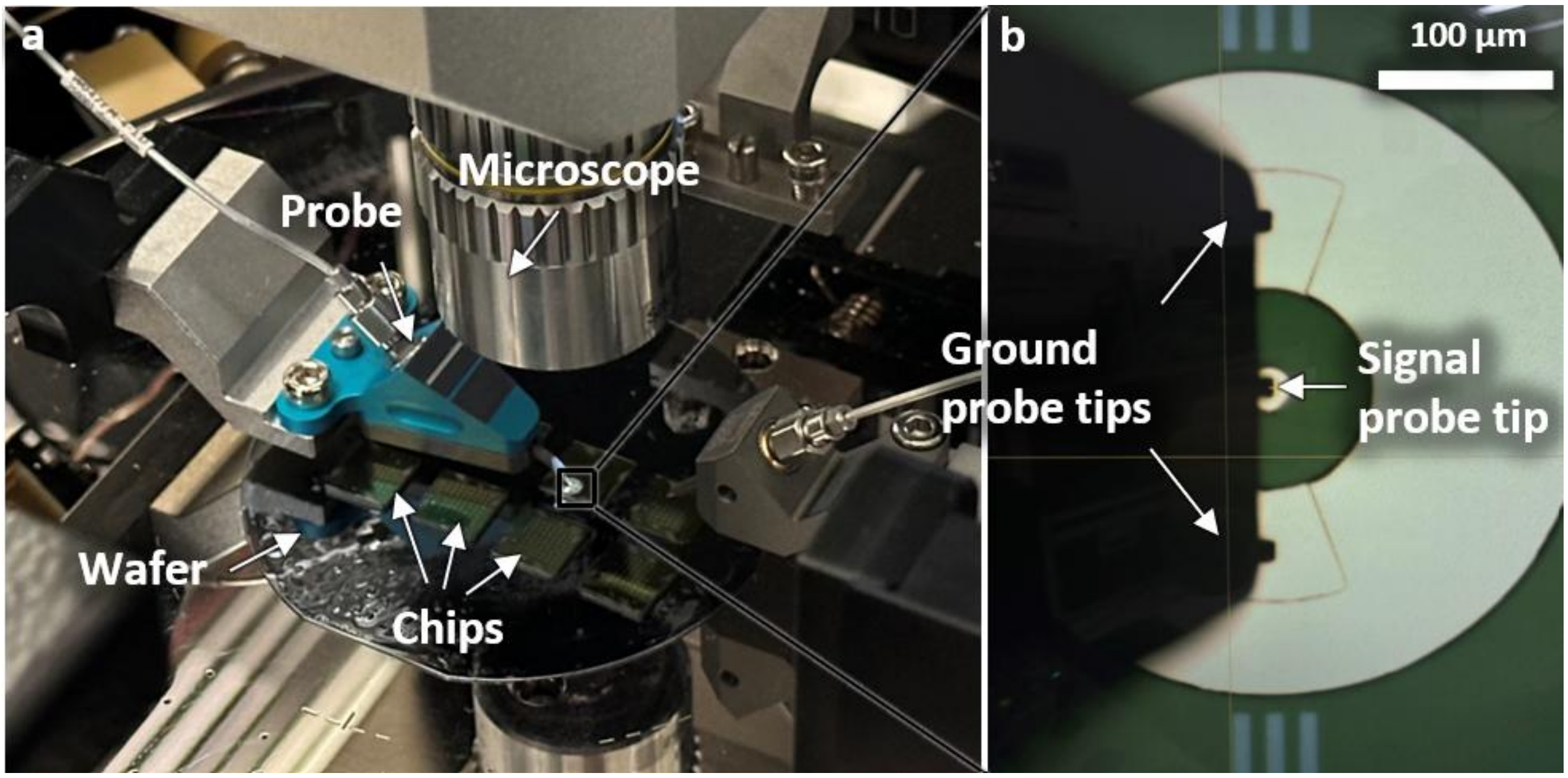

**Figure S14 |** Microwave measurement setup. **a**, Mounted carrier wafer with glued chips on top. **b**, Microscope image of the landed on-wafer probe on a device.

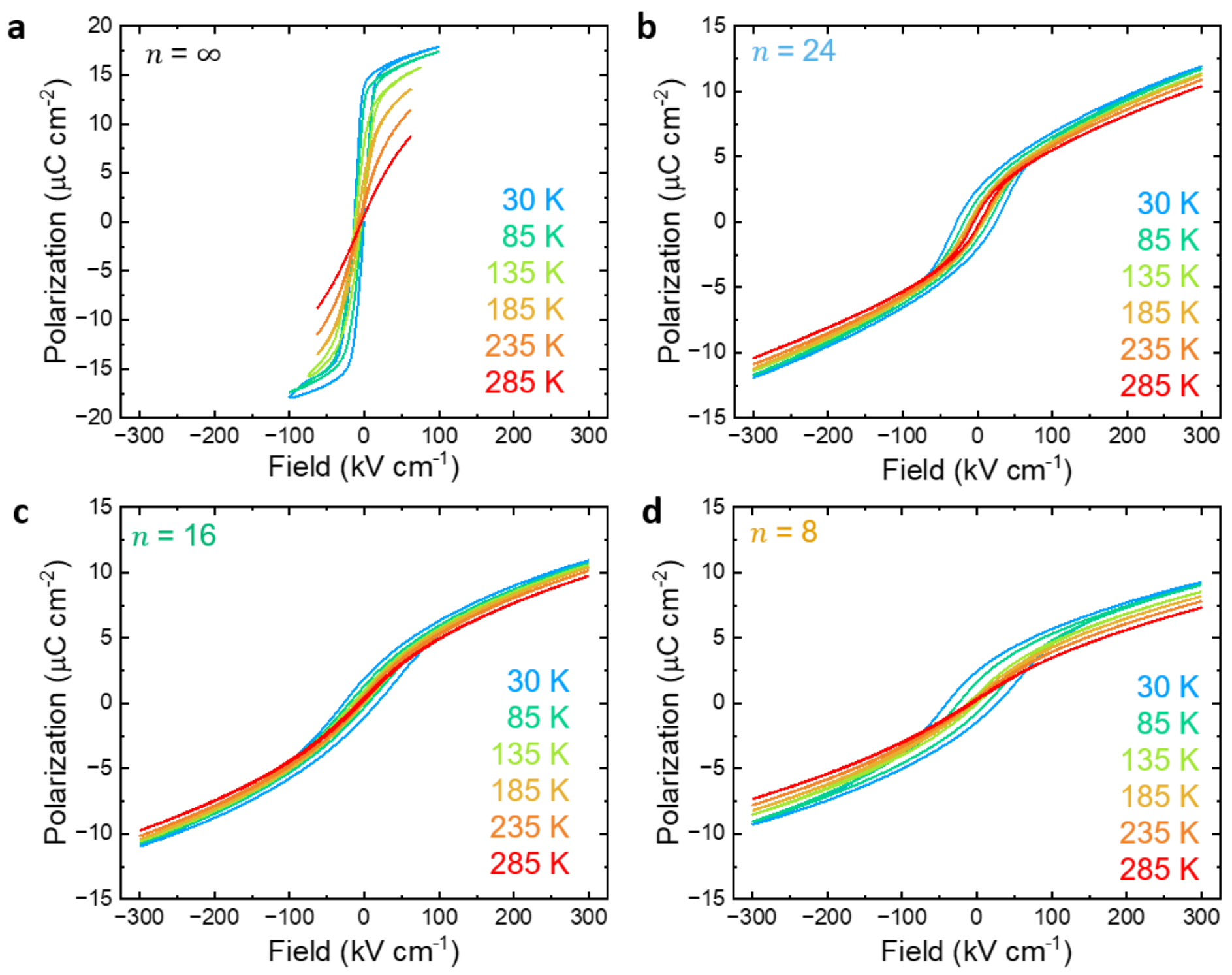

**Figure S15 |** Out-of-plane Polarization-Electric field loops of the four RP $(Ba_{0.45}Sr_{0.55}TiO_3)_nSrO$ ($n$ = 8, 16, 24, and ∞) thin films at various temperatures.

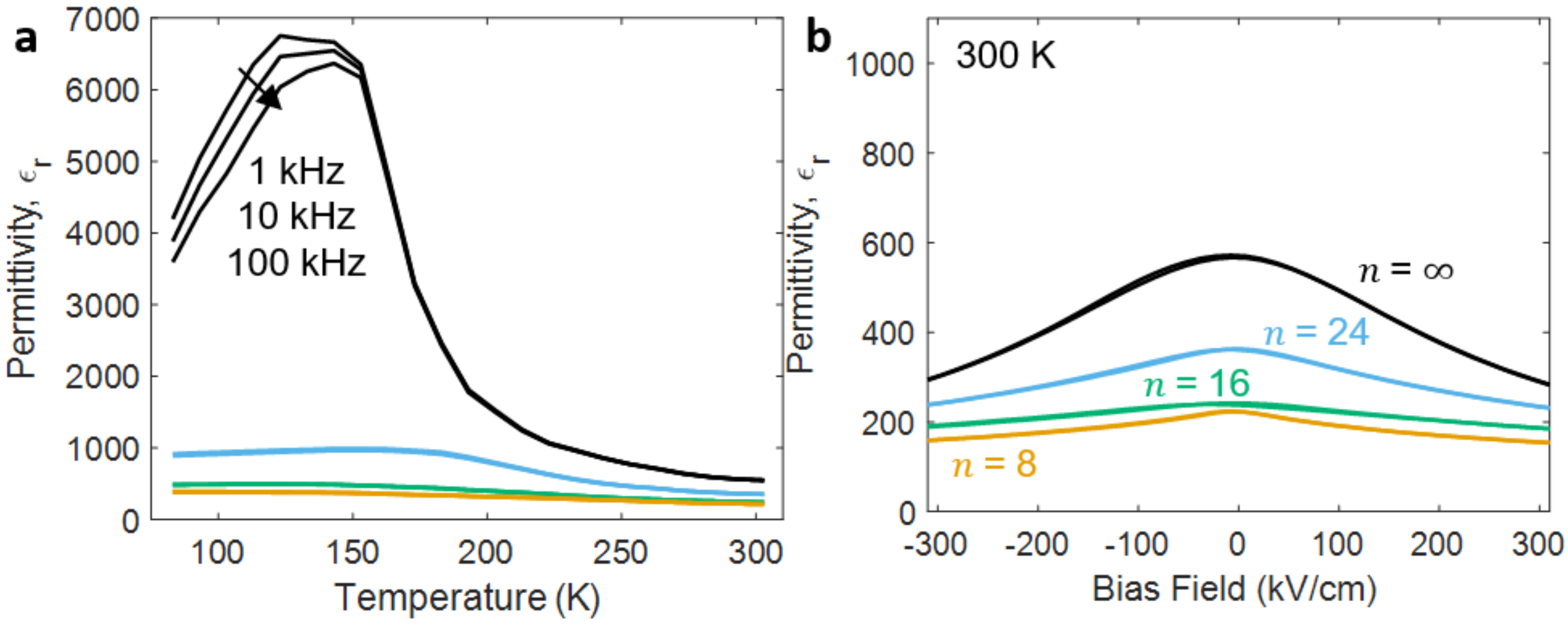

**Figure S16 |** Low-frequency permittivity of 4 RP $(Ba_{0.15}Sr_{0.85}TiO_3)_nSrO$ ($n$ = 8, 16, 24, and ∞) as a function of **a** temperature and b DC bias electric field at ambient temperature.

## Microwave characterization

For the MIM capacitors, we cycled through a DC bias voltage of ±5 V with a step size of at most 0.5 V. Before measuring a full cycle of 0 V, ..., -5 V, ..., 0 V, ..., 5 V, ..., 0 V, we added an additional half cycle to bias the film to account for potential ferroelectric hysteresis, which we did not observe. We presented permittivity data from 10 MHz up to 30 GHz. Below 10 MHz, the device impedance of the capacitors was too large to resolve in the $50\ \Omega$ reference impedance system of the network analyzer. Above 30 GHz, the data became unreliable because of two reasons. First, we extracted a nonphysical dispersive bottom electrode conductivity (**Fig. S19**). We hypothesize that this dispersion stems from the field pattern in the device becoming non-radial due to the finite conductivity of the top electrode[30]. Unfortunately, our permittivity extraction model relies on a radial symmetry of the electromagnetic field. Second, the differentiation of the impedance of the MIM capacitors from the impedance of the surface-shorts becomes sensitive to the repeatability of the electrical contact between the on-wafer probe and the test devices[30]. These two reasons limited our ability to characterize the permittivity above 30 GHz.

We evaluated the through-shorts to extract the bottom electrode conductivity. We observed different conductivities for the 4 RP chips. While most through-shorts on the same chip resulted in a very similar conductivity, some through-shorts resulted in lower conductivities (**Fig. S19**). We attribute these differences to inhomogeneous bottom electrodes or imperfect electrical contact between the gold top electrode and the $SrRuO_3$ bottom electrode. From the through-shorts that returned similar conductivities on one chip, we calculated a mean bottom electrode conductivity for each of the 4 RP chips. Then, we used this mean bottom electrode conductivity to extract the complex permittivity of the RP films. Using this mean bottom electrode conductivity, some capacitors resulted in an unphysical, negative loss for the dielectric film. Again, we suspect that this negative loss is due to a higher bottom electrode conductivity than we extracted from the through-shorts. This higher conductivity could stem from the inhomogeneity or imperfect electrical contact between the top electrode and the bottom electrode, like the different conductivities that we extracted from the through-shorts. We excluded these devices with significant negative loss from the evaluation.

In general, the Havriliak-Negami model satisfactorily describes the dielectric dispersions extracted from all MIM capacitors. Only for capacitors with a large permittivity above 10 GHz we observed systematic deviations between the data and the model (**Figs. 4a,b and S20**). We attribute this deviation to an imperfect landing repeatability of the on-wafer probe tip on the MIM capacitors and the calibration devices[30]. This uncertainty in repeatability becomes non-negligible when the impedance of the capacitors is low, which is the case for high permittivity and high frequencies. The imperfect contact repeatability appeared as apparent frequency dispersion in the real permittivity that is not accompanied with a corresponding increase in the imaginary permittivity. The deviation was most apparent for the combination of high frequency and high permittivity in the $n$ = 16 film.

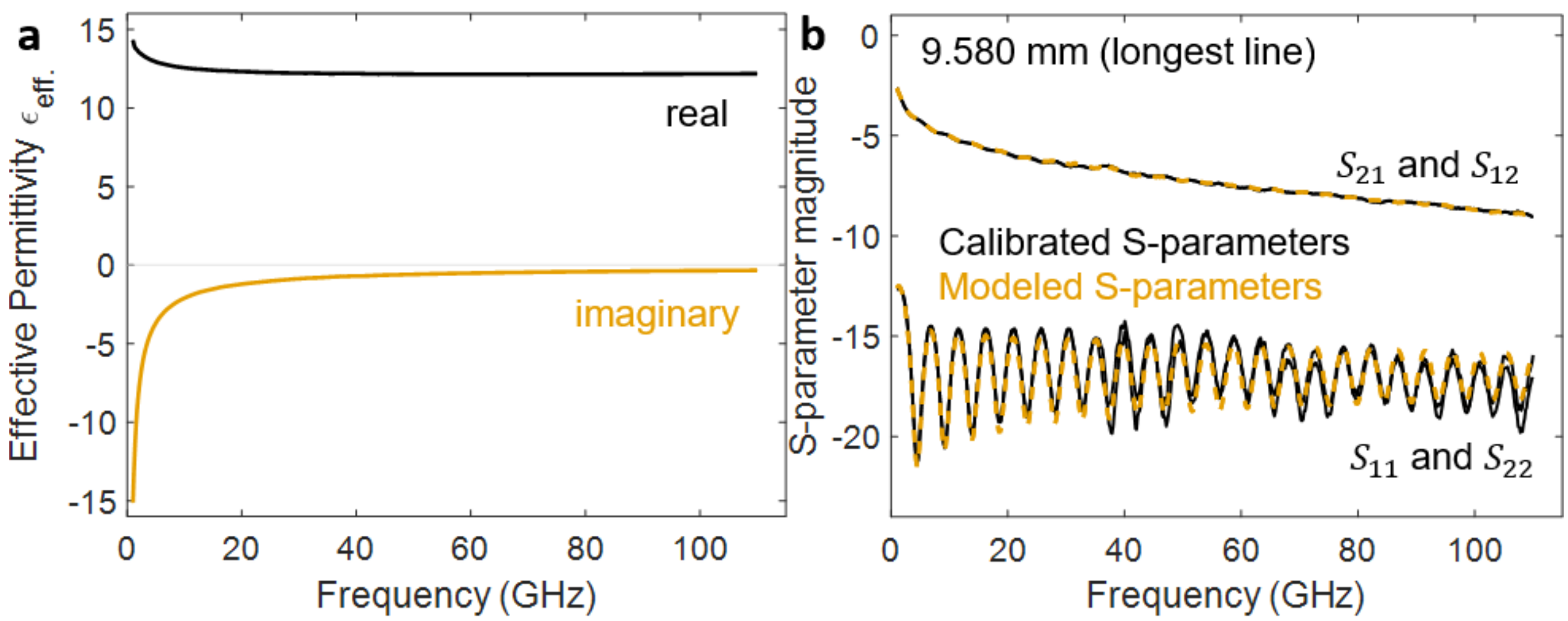

**Figure S17 |** Control of the 1st-tier mTRL calibration on the LSAT substrate. **a**, Effective permittivity of the CPW transmission lines and **b**, comparison of calibrated and modeled S-parameters for the longest line.

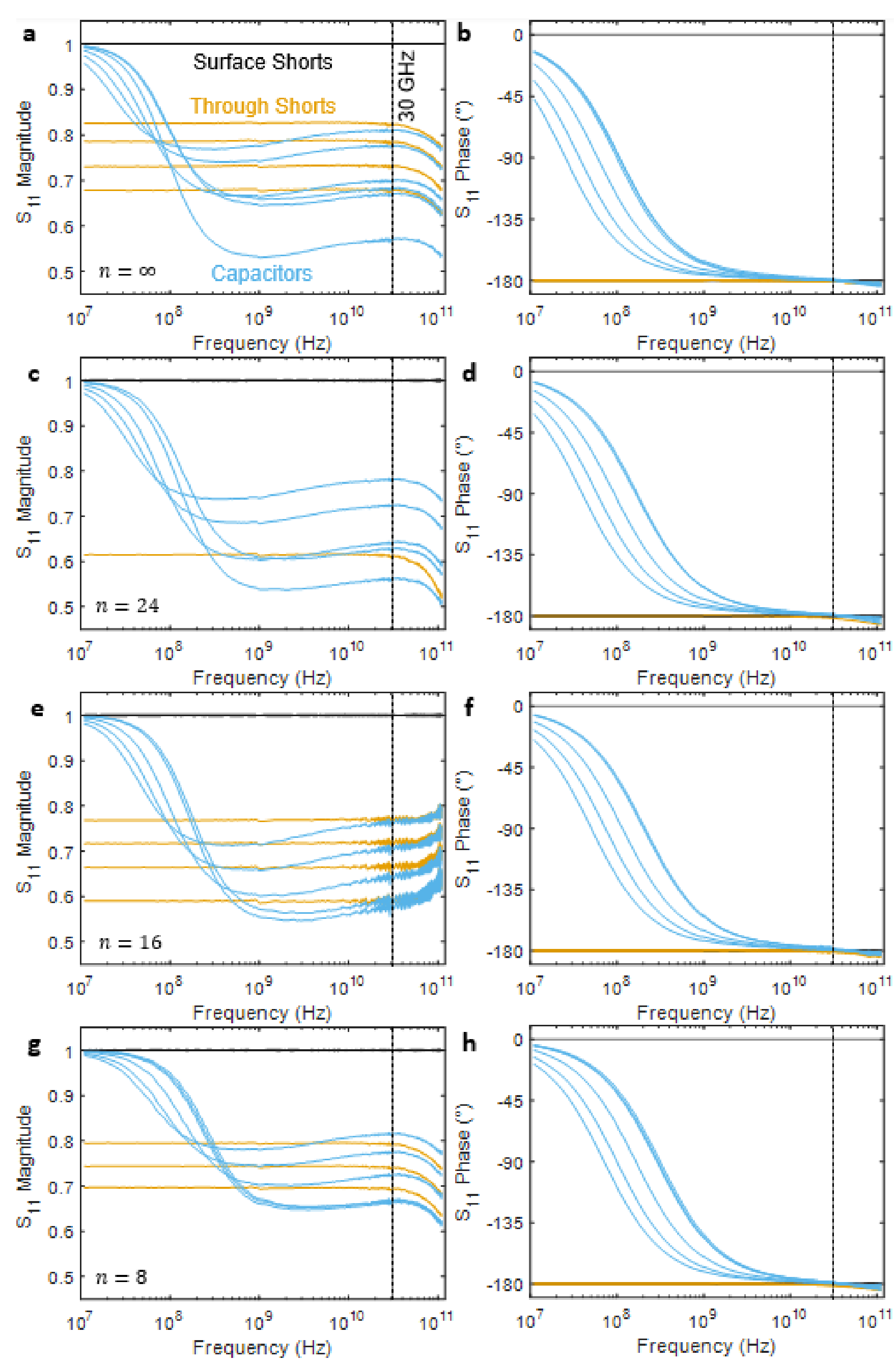

**Figure S18 |** 2nd-tier calibrated S-parameters of well-behaved surface-short, thru-short and MIM capacitor devices. Each line stands for an individual device with different geometry. The vertical line marks the maximum frequency (30 GHz) that we extract our out-of-plane permittivity up to.

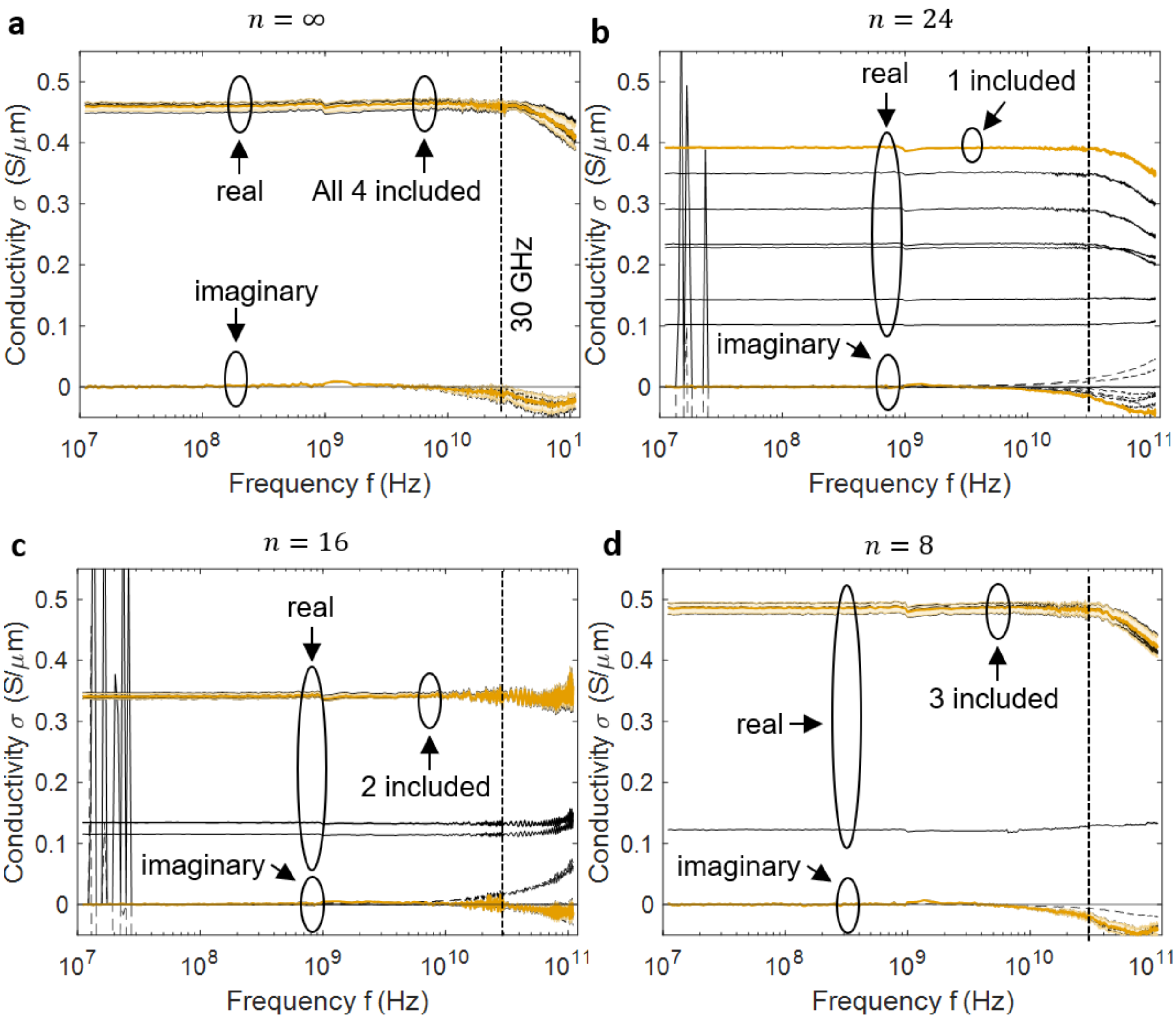

**Figure S19 |** Conductivity of the $SrRuO_3$ bottom electrode extracted from the thru-shorts. The black lines show the extracted conductivity from well-behaved (*i.e.*, highest conductivity) and ill-behaved (*i.e.*, lower conductivity) thru-short devices. The orange lines show the average of the conductivity extracted from well-behaved thru-short devices as its mean and its standard deviation.

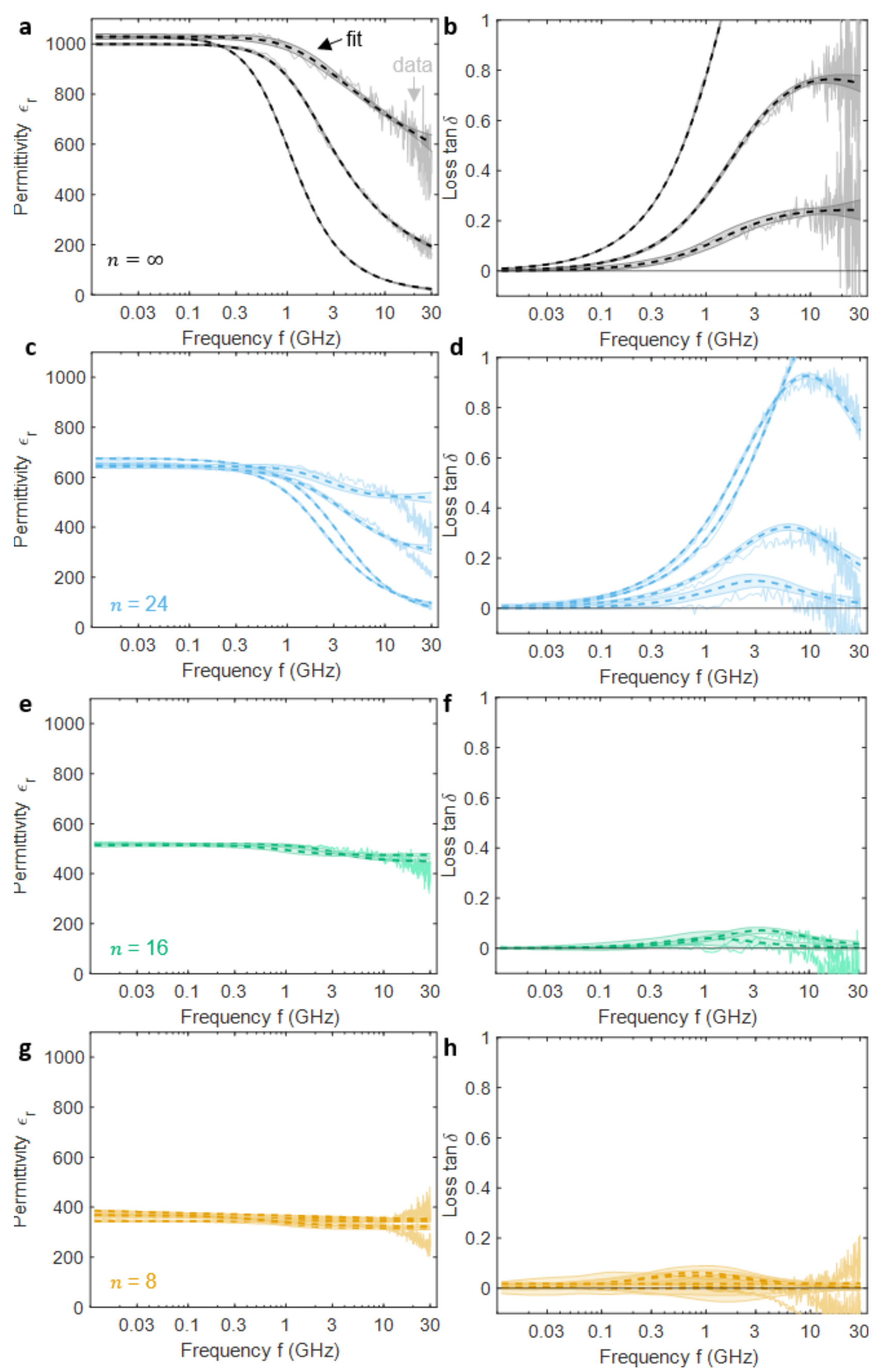

**Figure S20 |** Extracted complex out-of-plane permittivity for the 4 RP films from MIM capacitor devices with no significant extracted negative dielectric loss at different locations on each RP film. Real permittivity (**a**,**c**,**e**,**g**) and loss tangent (**b**,**d**,**f**,**h**). Each pair of data and fit originate from a different capacitor on the respective chip.

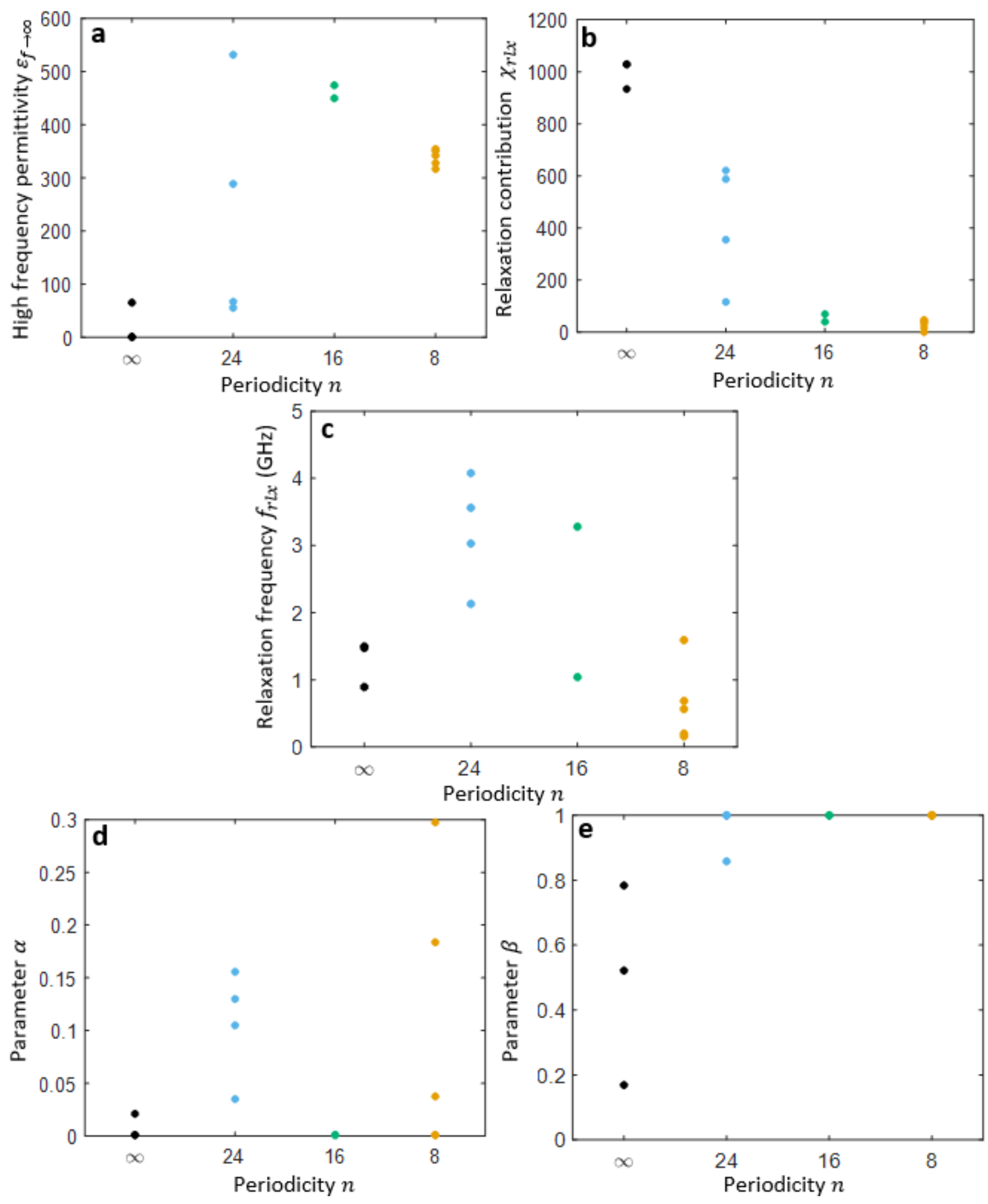

**Figure S21 |** Havriliak-Negami fit parameters (**Eq. 6**) of the MIM capacitors with no significant extracted negative dielectric loss.

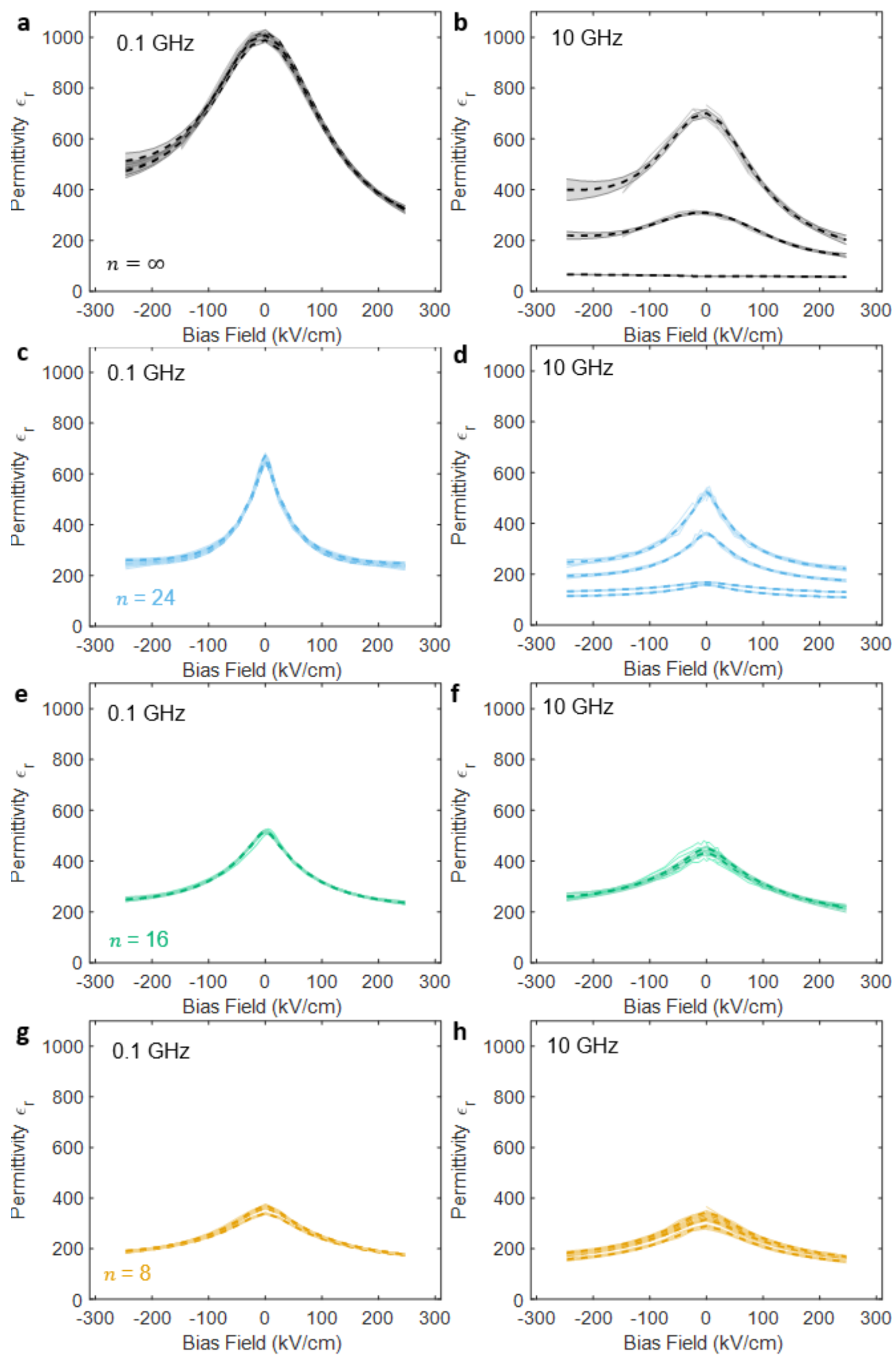

**Figure S22 |** Microwave tunability for all MIM capacitor devices with no significant extracted negative dielectric loss at 0.1 GHz (a,c,e,g) and 10 GHz (b,d,f,g).

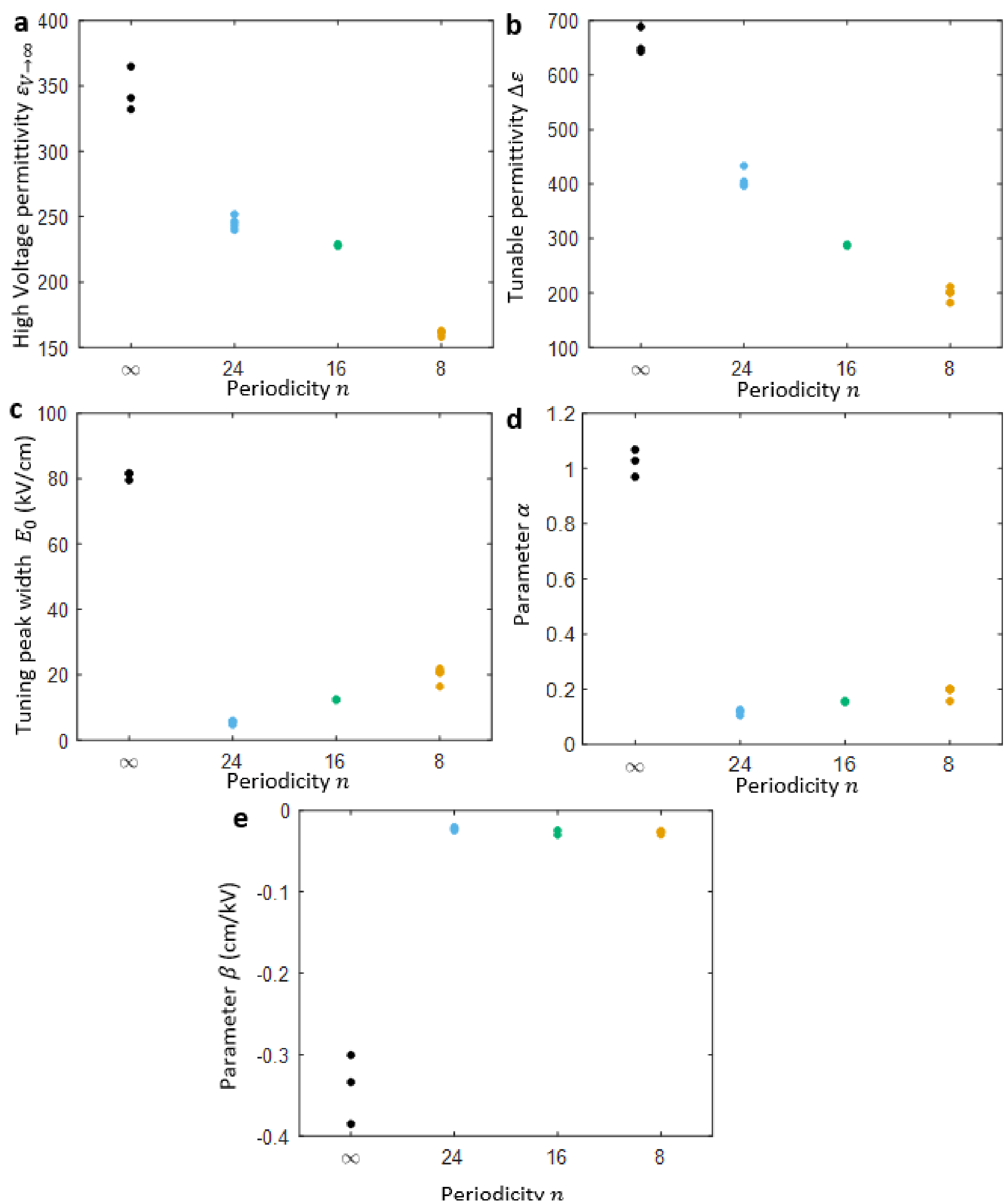

**Figure S23 |** Fit parameters of the real permittivity as a function of voltage at 0.1 GHz, based on the modified double-well potential model (**Eq. 7**).

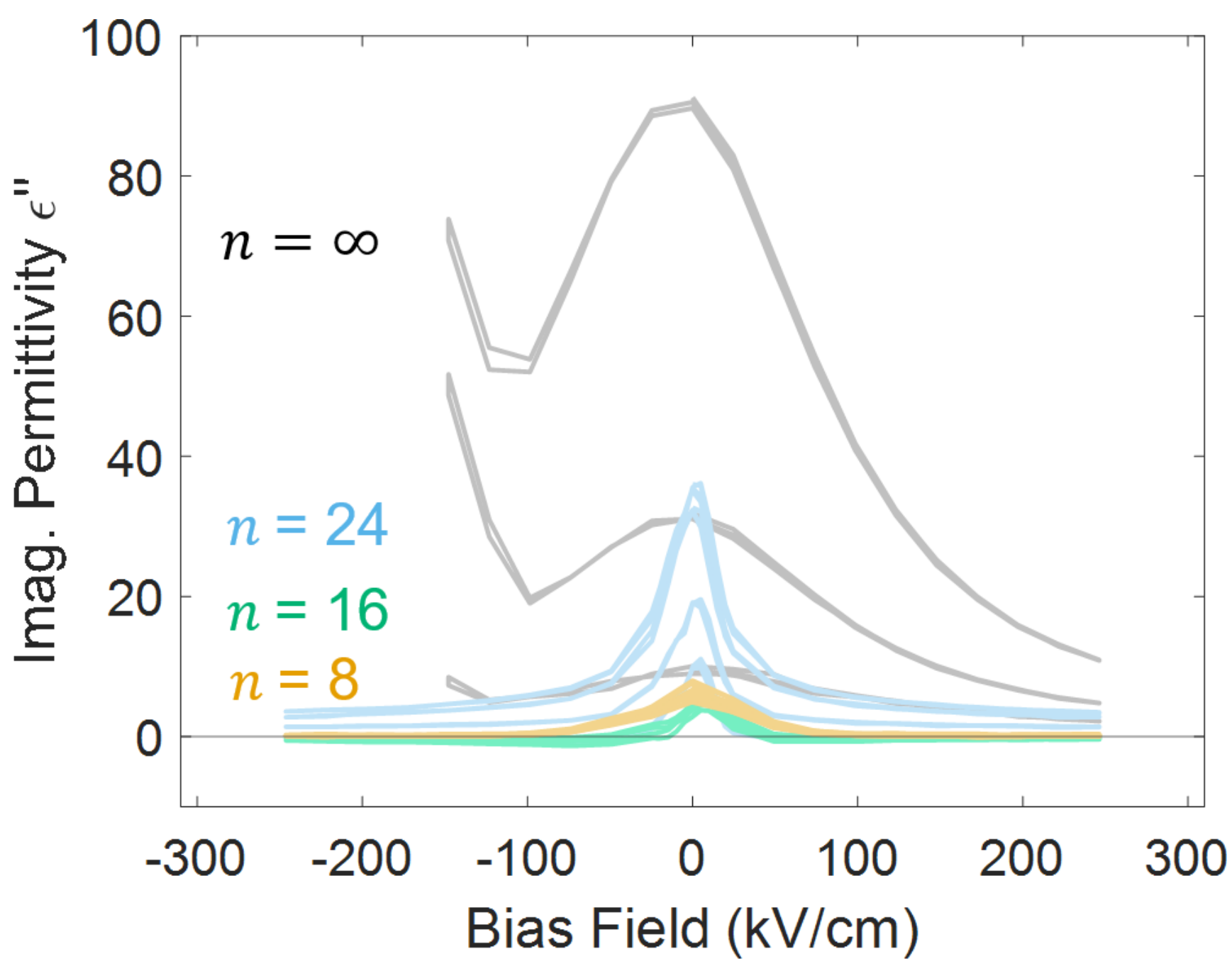

**Figure S24 |** Imaginary part of the relative permittivity extracted from the well-behaved MIM capacitors at 0.1 GHz.

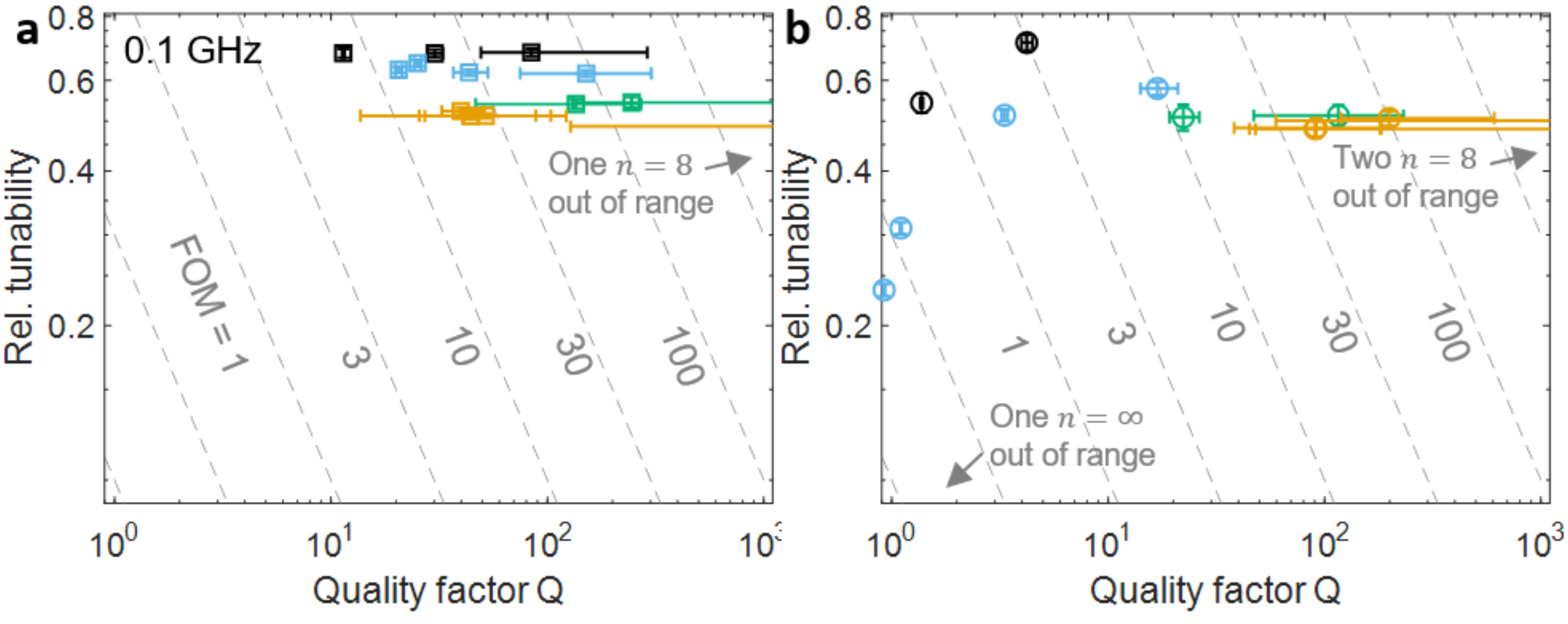

**Figure S25 |** Relative out-of-plane tunability, material quality factor and figure of merit (FoM) for all MIM capacitors with no significant extracted negative dielectric loss. **a**, At 0.1 GHz and **b**, at 10 GHz.